\documentclass[twocolumn,showpacs,preprintnumbers,amsmath,amssymb,prb,superscriptaddress]{revtex4}

\usepackage{graphicx}
\usepackage{dcolumn}
\usepackage{bm}
\usepackage{amsmath,amssymb}

\newcommand{\be}{\begin{equation}}
\newcommand{\ee}{\end{equation}}

\newcommand{\beq}{\begin{eqnarray}}
\newcommand{\eeq}{\end{eqnarray}}

\def\eq#1{(\ref{#1})}
\def\H1{\widehat{H}_1}

\begin{document}

\title{Screening in the two-dimensional electron gas with 
 spin-orbit coupling}

\author{M. Pletyukhov}
\affiliation{Institut f\"ur Theoretische Festk\"orperphysik, Universit\"at
Karlsruhe, D-76128 Karlsruhe, Germany}

\author{V. Gritsev}
\affiliation{D\'epartement de Physique, Universit\'e de Fribourg,
CH-1700 Fribourg, Switzerland}
\affiliation{Department of Physics, Harvard University, Cambridge, 
Massachusetts 02138, USA}

\begin{abstract}
We study screening properties of the two-dimensional electron gas
with Rashba spin-orbit coupling. Calculating the dielectric function
within the random phase approximation, we describe the new features
of screening induced by spin-orbit coupling, which are the extension
of the region of particle-hole excitations and the
spin-orbit-induced suppression of collective modes. The required
polarization operator is calculated in an analytic form without any
approximations. Carefully deriving its static limit, we prove the
absence of a small-$q$ anomaly at zero frequency. On the basis of
our results at finite frequencies we establish  the new boundaries
of the particle-hole continuum and calculate the SO-induced lifetime
of collective modes such as plasmons and longitudinal optical
phonons. According to our estimates, these effects can be resolved
in inelastic Raman scattering. We evaluate the experimentally
measurable dynamic structure factor and establish the range of
parameters where the described phenomena are mostly pronounced.
\end{abstract}
\maketitle

\section{Introduction}

Fundamental issues of interaction effects in the two-dimensional electron gas
(2DEG) are at the center of discussions since the early days of its fabrication
 \cite{AFS}. Quite generally, screening described by the dielectric function
forms the basis for understanding a variety of static and dynamic many-body
effects in electron systems \cite{Mahan}. The dielectric function of the
2DEG was computed a long time ago by Stern \cite{Stern} within the
random phase approximation (RPA). Different quasiparticle and collective
(plasma) properties deduced from those expressions were confirmed
experimentally soon after \cite{ATS}. Recent experiments measuring plasmon
dispersion, retardation effects, and damping \cite{Nagao,Kuk,West}
unambiguously show the importance of correlations between electrons.

More recently, the possibility of manipulating spin in 2DEG by
nonmagnetic means has generated a lot of activity \cite{ZFS}.
The key ingredient is  the Rashba spin-orbit (SO) coupling\cite{BR}
tunable by an applied electric field \cite{Nitta}.  A recent example
of SO-induced phenomena which has attracted much attention is the spin-Hall
effect \cite{SHE}.

In this context the study of interplay between electron-electron
correlations and SO coupling in 2DEG becomes an important problem.
In the preceding papers it has been already discussed how SO coupling
affects static screening\cite{Raikh}, plasmon dispersion and its
attenuation\cite{MCE,MH,wang}, and Fermi-liquid parameters\cite{SL}.
However, the approaches of these papers as well as of many
other papers on Rashba spin-orbit coupling  are based on various
approximations. We can outline the most popular ones: a linearization of
spectrum, also known as $\xi$-approximation; an expansion in SO coupling
parameter up to the lowest nonvanishing contribution; reshuffling the order
of momentum integration and evaluation of the zero-frequency limit in the 
polarization operator; a combination of any of those approximations.
In our opinion, their accuracy is not comprehensively discussed. Since its
lack might seriously affect a description of SO-induced phenomena, this
issue should be thoroughly investigated.

In the present paper, we study the effects of the dynamic screening
in 2DEG with Rashba SO coupling described by the RPA dielectric
function in the whole range of momenta and frequencies.
The main part of our paper is devoted to the analytic evaluation of the
polarization operator, which does not employ any approximations.
The knowledge of the dielectric function allows us to predict new features
in the directly observable dynamic structure factor. In particular, we
observe a SO-induced extension of the particle-hole excitation region and
calculate the SO-induced broadening of the collective
excitations such as plasmons and longitudinal optical (LO) phonons.
We  obtain the values of lifetimes that lie in the range of
parameters experimentally accessible by now in the inelastic Raman scattering
measurements.

On the basis of our analytic results, we also revisit the earlier approaches,
estimating their accuracy and establishing the limits of their applicability.
Although the approximations mentioned above usually work well for a
conventional 2DEG (without SO coupling), we demonstrate that in case of the
Rashba spectrum they should be applied with caution. In our paper we discuss
the subtle features of the  2DEG with a SO coupling warning about this. 
Special attention is focused on a derivation of a zero-frequency (static)
limit of the polarization operator. We present an analytic result, which,
however, does not contain an anomaly at small momenta predicted in
Ref.~\onlinecite{Raikh}.

The paper is organized as follows. In Sec.~\ref{defin} we introduce the 
main definitions and notations. In Sec.~\ref{operator} we present a
detailed analytic calculation of the polarization operator $\Pi
(q,\omega)$ and discuss the most important modifications generated
by SO coupling. We also derive an asymptotic value of ${\rm Im} \Pi
(q, \omega)$ at small $q$ and compare it to the approximate
expressions known before\cite{MCE,MH}. In Sec.~\ref{static} we
implement a thorough analytic derivation of the static limit
$\lim_{\omega \to 0}  \Pi (q, \omega)$ and make a conclusion about
the absence of an anomaly at small momenta. In Sec.~\ref{structure}
we study the directly observable dynamic structure factor and
evaluate SO-induced plasmon broadening. The latter effect also
causes a modification of the energy-loss function that is estimated
as well. In Sec.~\ref{phonons} we calculate the lifetime of LO
phonons generated by SO coupling.

\section{Basic Definitions}
\label{defin}
We consider a 2DEG with SO coupling of the Rashba type \cite{BR}
described by the single-particle Hamiltonian,
\be
H =\frac{{\bf k}^{2}}{2m^{*}}+\alpha_{R}{\bf n}(\mbox{\boldmath$\sigma$}
\times {\bf k}),
\label{initham}
\ee
where  ${\bf n}$ is a unit vector normal to the plane of 2DEG and
$\hbar=1$. The dispersion relation is SO split into two subbands
labeled by $\mu =\pm$,
\be
\epsilon_{{\bf k}}^{\mu} =\frac{k^2}{2 m^*} + \mu \alpha_{\mathrm{R}}k \equiv
\frac{(k +\mu k_R)^2}{2 m^*} -  \frac{k_R^2}{2 m^*}.
\label{SOspectrum}
\ee
These subbands have the distinct Fermi momenta $k_{\mu} = k_F -\mu k_R$ and
the same Fermi velocity $v_F = k_{F} /m^*$.
Here we denote $k_R = m^* \alpha_R$ and $k_F = \sqrt{2 m^* E_F + k_R^2}$, 
and assume that $2 k_R < k_F$.

The effective Coulomb interaction  is
$V^{eff}_{q \omega}=V_q /\varepsilon_{q \omega}$, where
$V_q=2\pi e^2/(q\varepsilon_{\infty})$, and $\varepsilon_{\infty}$ is the
(high-frequency) dielectric constant of medium. The dielectric function
$\varepsilon_{q \omega} = {\rm Re} \, \varepsilon_{q \omega}+ i {\rm Im} \,
\varepsilon_{q \omega}$ describes effects of dynamic screening, and in
the random phase approximation (RPA) it is given by \cite{Mahan}
\be
\varepsilon_{q \omega} = 1-V_q \Pi_{q\omega},
\label{rpa}
\ee
where $\Pi_{q \omega}$ is a polarization operator. In the presence of
SO coupling, the latter is a sum
\be
\Pi_{q \omega} = \sum_{\chi,\mu=\pm}\Pi_{q \omega}^{\chi,\mu}
\label{bub}
\ee
of contributions
\be
\Pi^{\chi,\mu}_{q \omega} = \lim_{\delta \to 0}
\int \frac{d^2 {\bf k}}{(2 \pi)^2}\,\,
\frac{n_F (\epsilon_{{\bf k}}^{\mu}) 
-n_F (\epsilon_{{\bf k}+{\bf q}}^{\chi \mu})}
{\omega+ i \delta 
+ \epsilon_{{\bf k}}^{\mu} - \epsilon_{{\bf k}+{\bf q}}^{\chi \mu}}
{\cal F}^{\chi}_{{\bf k}, {\bf k}+{\bf q}},
\label{basicdef2}
\ee
where the indices $\chi=+$ and $\chi=-$ correspond to the intersubband and
intrasubband transitions, respectively. The form factors
\be
{{\cal F}}^{\chi}_{{\bf k}, {\bf k}+{\bf q}} =
\frac12 [1 + \chi \cos (\phi_{{\bf k}} - \phi_{{\bf k}+{\bf q}})]
\ee
originate from the rotation to the eigenvector basis, and
\beq
\cos (\phi_{{\bf k}} - \phi_{{\bf k} + {\bf q}}) =\frac{|{\bf k}| +
x |{\bf q}|}{|{\bf k} + {\bf q}|}, \nonumber \\
x = \cos (\phi_{{\bf k}} - \phi_{{\bf q}}) \equiv \cos \phi.
\label{angul}
\eeq

Throughout the paper we will use the dimensionless units $y= k_R /
k_F$, $z=q / 2 k_F$, $v= k/ k_F$, and $w=m^* \omega/2 k_F^2$.

\section{Polarization operator at arbitrary frequency}
\label{operator}
\subsection{Evaluation of the polarization operator}
For a calculation of \eq{basicdef2} it is useful to shift ${\bf k}
\to {\bf k} - {\bf q}$ in those terms of \eq{basicdef2} that
contain $n_F (\epsilon_{{\bf k}+{\bf q}}^{\chi \mu})$. At the same
time we can shift the integration angle $\phi \to \phi +\pi$ in the
same terms due to the momentum isotropy of the spectrum
\eq{SOspectrum}. Conveniently regrouping all contributions in
Eq.~\eq{bub}, we cast it into the form
\be
\Pi_{q \omega} = \sum_{\mu,\lambda =\pm} \Pi_{q \omega, \lambda}^{\mu} ,
\ee
where
\beq
& & \Pi_{q \omega, \lambda}^{\mu}= \lim_{\delta \to 0} 
\int \frac{d^2 {\bf k}}{(2 \pi)^2} n_F (\epsilon_{{\bf k}}^{\mu}) \\
& & \times \left[ \frac{{\cal F}^+_{{\bf k},{\bf k}+{\bf
q}}}{\epsilon_{{\bf k}}^{\mu} - \epsilon_{{\bf k}+{\bf q}}^{\mu} +
\lambda (\omega + i \delta)} +  \frac{{\cal F}^-_{{\bf k},{\bf
k}+{\bf q}}}{\epsilon_{{\bf k}}^{\mu} -  \epsilon_{{\bf k}+{\bf
q}}^{-\mu} + \lambda (\omega + i \delta)} \right], \nonumber
\eeq
and the index $\lambda$ effectively labels the contributions from
the different (``in'' and ``out'') Fermi functions.

In the limit of zero temperature we obtain
\beq
& & \Pi_{q \omega, \lambda}^{\mu} =
\frac{1}{8 \pi^2} \lim_{\delta \to 0}
\int_0^{k_F - \mu k_R} k d k \int_0^{2 \pi} d \phi
\label{pilm}\\
& & \times \left[ \frac{1 + \cos (\phi_{{\bf k}} - \phi_{{\bf k}+ {\bf q}})}
{\epsilon_{{\bf k}}^{\mu} - \epsilon_{{\bf k}+ {\bf q}}^{\mu} +
\lambda (\omega+ i \delta)} 
+ \frac{1 - \cos (\phi_{{\bf k}} - \phi_{{\bf k}+{\bf q}} )}
{\epsilon_{{\bf k}}^{\mu} - \epsilon_{{\bf k}+ {\bf q}}^{-\mu} + 
\lambda (\omega + i \delta)} \right]  .
\nonumber
\eeq
After the intermediate steps \eq{intermed1},
\eq{intermed2} we cast \eq{pilm} into the form
\beq
- \frac{1}{\nu} {\rm Im}\Pi_{q \omega, \lambda}^{\mu} &=& \int_0^{1
-\mu y} v g_i (v,z, w, y) d v ,\label{pregunf} \\
- \frac{1}{\nu} {\rm Re} \Pi_{q \omega, \lambda}^{\mu} &=& \int_0^{1
-\mu y} v f_i (v,z, w, y) d v ,
\label{prefunf}
\eeq
where $\nu \equiv \nu_{2D} = m^*/(2 \pi)$ is the density of states
in 2DEG per each spin component; the indices $i=1,2,3,4$ correspond
to $\{\mu, \lambda \} = \{ -,+\},\{ +,+\},\{ -,-\},\{ +,-\}$, and
the functions $g_i$, $f_i$ are defined by
\beq
g_i &=& \frac{\lambda C_i}{2} \int_{0}^{2 \pi} d \phi  \,\, 
{\rm sign} (2 v z x - \mu y v + 2 (z^2 -\lambda w)) \nonumber \\
& & \qquad \times  (x+ \delta_i) \,\, \delta (x^2 +\beta_i x +\gamma_i) ,
\label{gunf} \\
f_i &=& \frac{C_i}{2 \pi} \int_0^{2 \pi} d \phi
\frac{x + \delta_i}{x^2 + \beta_i x + \gamma_i},
\label{funf}
\eeq
with the coefficients
\beq
C_i &=& \frac{v -\mu y}{2 v^2 z}, \quad \delta_i =
\frac{(z^2 - \lambda w) -\mu y v}{(v -\mu y) z},  \nonumber \\
\beta_i &=& \frac{2 (z^2-\lambda w) - \mu y (v +\mu y)}{v z}, \nonumber \\
\gamma_i &=& 
\frac{(z^2 - \lambda w)^2 -\mu y v (z^2 - \lambda w) -z^2 y^2}{v^2 z^2}. 
\label{ingred}
\eeq
In \eq{gunf} the Dirac delta function is denoted by $\delta (\ldots)$,
as usual.

The equation $x^2 + \beta_i x + \gamma_i =0$ has the roots labeled 
by $\sigma = \pm$,
\be
\lambda_{i \sigma} = \frac{\mu y (v+ \mu y) - 2 (z^2- \lambda w) +
\sigma y \sqrt{(v+ \mu y)^2 + 4 \lambda w}}{2 v z} .
\label{roots}
\ee
The roots $\lambda_{1\sigma}$ and $\lambda_{2 \sigma}$ are always real, while
$\lambda_{3 \sigma}$ and  $\lambda_{4 \sigma}$ become complex in the ranges
$y- 2 \sqrt{w} < v < y+ 2\sqrt{w}$ and  $-y- 2 \sqrt{w} < v < -y+ 2\sqrt{w}$, 
respectively.

All integrals in \eq{pregunf} and \eq{prefunf} receive a contribution from
those ranges of $v$ where
$\lambda_{i \sigma}$ are real. Additionally, the integrals $\int v f_{3,4} dv$
receive a contribution from the ranges with complex $\lambda_{3,4 \sigma}$.
In order to take into account both types of contribution, we represent the
function $f_i = f_i^I + f_i^{II}$ as a sum of $f_i^I = f_i \Theta
(\beta_i^2 - 4 \gamma_i)$ and $f_i^{II} = f_i \Theta (4 \gamma_i - \beta_i^2)$.
Obviously, $f_{1,2}^{II} \equiv 0$.

Let us also introduce ${\rm Re} \Pi^I$ and ${\rm Re} \Pi^{II}$ that are
obtained by integrating the functions $\sum_i v f_i^I$ and $\sum_i v f_i^{II}$,
respectively. Thus, the full real part of the polarization operator is given
by the sum of the two, i.e.,
\be
{\rm Re} \Pi = {\rm Re} \Pi^I + {\rm Re} \Pi^{II}.
\label{thesumpi}
\ee
Note that in the limit $w \to 0$ all the roots $\lambda_{i \sigma}$ 
become real,
and it might seem that ${\rm Re} \Pi^{II}$ vanishes in the static limit.
However, it is not the case, and ${\rm Re} \Pi^{II}$ does give 
a finite contribution
as $w \to 0$. We will discuss this point in a deep detail in the next section.

For $\beta_i^2 > 4 \gamma_i$ we have
\beq
\delta (x^2 + \beta_i x + \gamma_i) = \frac{\delta (x -\lambda_{i+}) 
+ \delta (x -\lambda_{i-})}{\lambda_{i+} - \lambda_{i-}}, \\
\frac{x + \delta_i}{x^2 + \beta_i x + \gamma_i} =
\frac{1}{\lambda_{i+} - \lambda_{i-}} \left( \frac{\lambda_{i+} +
\delta_i}{x - \lambda_{i+}} - \frac{\lambda_{i-} + \delta_i}{x -
\lambda_{i-}}\right).
\eeq
Using the integrals \eq{deltint} and \eq{angint} we establish that
$g_i = \sum_{\sigma} g_{i\sigma}$, $f_i^I = \sum_{\sigma} f_{i\sigma}^I$, and
\beq
& & g_{i \sigma} = \frac{\lambda C_i}{\lambda_{i+} - \lambda_{i-}}
\frac{\lambda_{i \sigma}+\delta_i}{\sqrt{1-\lambda_{i \sigma}^2}} 
\Theta (1-|\lambda_{i \sigma}|)  \\
& \times & \Theta ((v +\mu y)^2 + 4 \lambda w) \,\, 
{\rm sign} (y + \sigma \sqrt{(v +\mu y)^2 + 4 \lambda w}), \nonumber
\eeq
\be
f_{i\sigma}^I = - \frac{\sigma C_i}{\lambda_{i+} - \lambda_{i-}}
\frac{\lambda_{i\sigma} + \delta_i}{\sqrt{\lambda_{i \sigma}^2 - 1}}
\Theta (|\lambda_{i \sigma} | - 1) {\rm sign} (\lambda_{i \sigma} ) .
\label{fIsigma}
\ee

For $\beta_i^2 < 4 \gamma_i$ we have
\be
-\frac{1}{\nu} {\rm Re} \Pi^{II}
= \int_{{\cal D}_-} v f_{3}^{II} d v + \int_{{\cal D}_+} v f_{4}^{II} d v,
\label{repi2}
\ee
where ${\cal D}_{\mp}$ are defined in the following way:
\beq
y^2 > 4 w &:& \,\, {\cal D}_- = [y -2 \sqrt{w}, y+ 2 \sqrt{w}], \\
4 w > y^2 \cap 1 > 4 w &:& \,\, {\cal D}_- = [0, y+ 2 \sqrt{w}], \\
4 w >1 &:& \,\, {\cal D}_- = [0, 1+y],
\eeq
and
\beq
y^2 > 4 w &:& \,\,  {\cal D}_+ = \emptyset , \\
4 w > y^2 \cap 1 > 4 w &:& \,\, {\cal D}_+ =[0,-y+2 \sqrt{w}], \\
4 w >1 &:& \,\, {\cal D}_+ = [0, 1-y].
\eeq
Applying the table integral \eq{angintcomp}, we obtain
\beq
f_{3,4}^{II} (v,z,w,y) &=& \frac{1}{2z \sqrt{(v_1^{\mu}-v)
(v-v_2^{\mu})}} \label{f3comp} \\
&\times &\frac{\sqrt{P_{\mu} (v)}+\sqrt{Q_{\mu}
(v)}}{\sqrt{\left(\sqrt{P_{\mu} (v)}+\sqrt{Q_{\mu} (v)} \right)^2 -
4 y^2 z}} \nonumber,
\eeq
where
\beq
P_{\mu} (v) &=& (z+\mu y) (v- v_2^{\mu}) (v_3^{\mu}-v), \\
Q_{\mu} (v) &=& (z-\mu y) (v- v_1^{\mu}) (v_4^{\mu}-v), \\
v_{1,2}^{\mu}&=& -\mu y \pm (z+w/z), \\
\quad v_{3,4}^{\mu} &=& \pm z + \frac{w}{\mu y \pm z},
\eeq
and the indices $\mu = -$ and  $\mu = +$ correspond to $f_3^{II}$ 
and $f_4^{II}$, respectively.

The numerical evaluation of \eq{repi2} can be performed with a controlled and
sufficiently high accuracy, since the functions \eq{f3comp} may diverge only
near the actual integration edges $-\mu y \pm 2 \sqrt{w}$. On the other hand,
one can try to find an analytic expression for  \eq{repi2}. An alternative
representation of $f_{3,4}^{II}$, which is equivalent to \eq{f3comp} and 
more suitable for further analytic evaluation, is introduced in
Appendix~\ref{compint}.

As for ${\rm Im} \Pi = -\nu \sum_{i=1}^4 \int_0^{1-\mu y} v g_{i} dv$ and
${\rm Re} \Pi^I = -\nu \sum_{i=1}^4 \int_0^{1-\mu y} v f_{i}^I dv$, their
analytic evaluation is presented below. Making a change of variables,
\be
\tau = \frac12 \left[-  \mu  (v+ \mu y) + \sigma \sqrt{(v+ \mu y)^2
+ 4 \lambda w}\right],
\label{chanvar}
\ee
we establish the relations \eq{voft}-\eq{vci}, and thus deduce
\beq
- \frac{1}{\nu} {\rm Im} \Pi  &=& \sum_{\sigma,\mu} \sigma 
\int_{\tau_{\sigma +} (y)}^{\tau_{\sigma +} (\mu)} d \tau {\cal L}^+ (\tau) 
\label{impi} \\
&-& \Theta (1 -  4 w) \sum_{\mu} \int_{\tau_{+-} (\mu)}^{\tau_{--} (\mu)} 
d \tau {\cal L}^- (\tau) \nonumber \\
&+& 2 \Theta (y^2 - 4 w) \int_{\tau_{--}(y)}^{\tau_{+-} (y)} 
d \tau {\cal L}^- (\tau) , \nonumber
\eeq
and
\beq
- \frac{1}{\nu} {\rm Re} \Pi^I &=&  \sum_{\sigma,\mu} 
\int_{\tau_{\sigma +} (y)}^{\tau_{\sigma +} (\mu)} d \tau {\cal R}^+ (\tau) 
\label{realpi} \\
&+& \Theta (1 -  4 w) \sum_{\sigma, \mu} 
\int_{- \mu \tau_{++} (0)}^{\tau_{\sigma -} (\mu)} d \tau {\cal R}^- (\tau) 
\nonumber \\
&+& 2 \Theta (y^2 - 4 w) \sum_{\sigma} 
\int_{\tau_{\sigma -}(y)}^{\tau_{-+} (0)} d \tau {\cal R}^- (\tau) , 
\nonumber
\eeq
where
\beq
\tau_{1,2} &=& \pm w/z, \quad \tau_{3,4} = -y \pm z ,
\label{tauk} \\
\tau_{\sigma \lambda} (x) &=& 
\frac12 \left[-x +\sigma \sqrt{x^2 + 4 \lambda w} \right], \\
{\cal L}^{\pm} (\tau) &=& {\cal L} (\tau) \,\, 
{\rm sign} (\tau^2 + y \tau \pm w), \\
{\cal R}^{\pm} (\tau) &=& {\cal R} (\tau) \,\, 
{\rm sign} (\tau z \mp w (\tau +y)/z), \\
{\cal L} (\tau)&=&\frac{1}{2 z} 
\frac{(\tau -\tau_3)(\tau -\tau_4)}{\sqrt{\prod_{k=1}^4
(\tau - \tau_k)}} \Theta \left(\prod_{k=1}^4 (\tau - \tau_k)\right), \\
{\cal R} (\tau)\!&=&\!\frac{1}{2 z} \frac{(\tau -\tau_3)(\tau
-\tau_4)}{\sqrt{-\prod_{k=1}^4 (\tau - \tau_k)}} \Theta\!\left(\!-\!
\prod_{k=1}^4 (\tau - \tau_k)\!\!\right)
\eeq

\begin{figure}[t]
\includegraphics[width=8.4cm,angle=0]{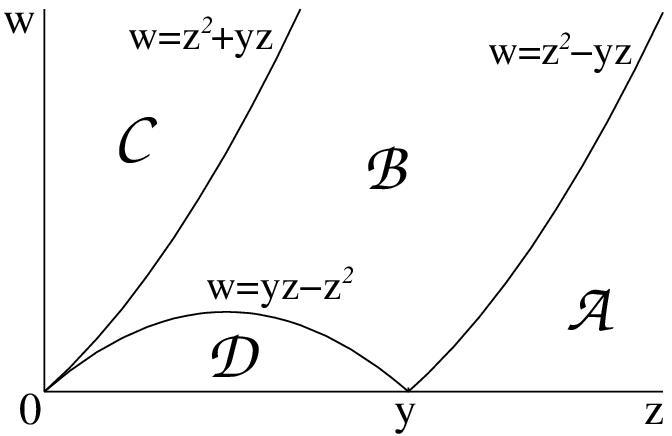}
\caption{ The domains ${\cal A}$, ${\cal B}$, ${\cal C}$, ${\cal D}$
[see Eqs.\eq{orda}-\eq{ordd}] corresponding to the
 different orderings of the roots \eq{tauk}.}
\label{domains}
\end{figure}

In Fig.~\ref{domains}, the quarter-plane $(z>0,w>0)$ is divided 
into the domains ${\cal A} = \{ (z,w)|w< z (z-y) \}$, 
${\cal B} = \{(z,w)|w>z (z-y)\cap w< z (z+y) \cap w> z (y-z)\}$,
${\cal C} = \{(z,w) |w>z (z+y) \}$, ${\cal D} =  \{(z,w)| w < z (y-z)\}$,
which are specified by an ordering of the roots $\tau_k$
\eq{tauk}:
\beq
{\cal A} &:& \quad \tau_4 < \tau_2 < \tau_1 <\tau_3, \label{orda}\\
{\cal B} &:& \quad \tau_4 < \tau_2 < \tau_3 < \tau_1, \label{ordb}\\
{\cal C} &:& \quad  \tau_2 < \tau_4 < \tau_3 < \tau_1 , \label{ordc} \\
{\cal D} &:& \quad \tau_4 < \tau_3< \tau_2 < \tau_1 \label{ordd}.
\eeq
In each domain one should compare $\tau_k$ and $\tau_{\sigma \lambda} (x)$ 
in order to establish actual limits of integration in \eq{impi} 
and \eq{realpi}. After that, it becomes possible to write down 
${\rm Im} \Pi$ and ${\rm Re} \Pi^I$ in an explicit form.
We refer to the Appendix \ref{elliptic} where the necessary expressions 
for establishing the explicit
form of \eq{impi} are presented. Similar expressions for \eq{realpi} 
can be  easily derived from Ref.~\onlinecite{gradst}.

\subsection{SO-induced extension of the particle-hole excitation region}

\begin{figure}[b]
\includegraphics[width=8.6cm,angle=0]{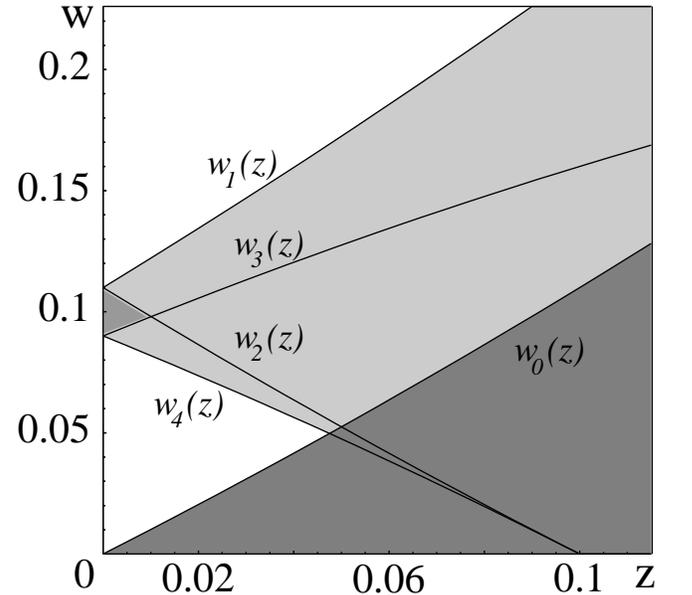}
\caption{SO-induced extension of the particle-hole continuum shown
for $y=0.1$. It is bounded from above by parabola $w_{1}(z)$
[Eq.~(\ref{omega1})] and bounded from below by the parabola
$w_{4}(z)$ [Eq.~(\ref{omega4})] and by the parabola $w_0
(z)=z^{2}+z$ of the conventional particle-hole (Landau damping)
region. The small darkened triangle indicates the region where the
approximation $\sigma (z,w) \approx \sigma (0,w)$ is applicable.}
\label{optrange}
\end{figure}

Let us analyze the most important modifications to the polarization 
operator induced by SO coupling. First of all, we are interested 
in establishing the new boundaries
of a  particle-hole continuum (or Landau damping region), 
which is defined by the condition
${\rm Im} \Pi \neq 0$.  They can be determined from a simple 
consideration of extremes of
the denominators in \eq{basicdef2}. On the basis of this 
purely kinematic argument, it is easy to
establish that due to SO coupling there
appears a new wedge-shaped region of damping 
(shown in Fig.~\ref{optrange}). It is bounded by the two parabolas 
$-(z-y)^2-(z-y) = w_4 (z) < w < w_1 (z) =(z+y)^2 + (z+y)$
and attached to the boundary $w_0 (z) =z^2+z$ of the conventional 
particle-hole continuum
(obtained in the absence of SO coupling according to Ref.~\onlinecite{Stern}).
Another boundary $w = z^2-z$ of the latter transforms into
$w = (z-y)^2 - (z-y)$ for nonzero $y$ (this occurs at $z>1$, 
and therefore it is not shown in Fig.~\ref{optrange}).

An extension of the particle-hole continuum reflects an opened
possibility for transitions between SO-split subbands. Therefore one
can expect new SO-induced effects of damping of various collective
excitations (plasmons, LO phonons) in the new regions of damping.
These issues will be discussed in the subsequent sections. Below we
present ${\rm Im} \Pi$  in explicit form for the values $w> w_0
(z)$, i.e., above the boundary $w_{0}(z)$ of the conventional
particle-hole continuum. Since $w_0 (z) > z^2 + y z$, we deal with
the case of the roots' ordering \eq{ordc} corresponding to the domain
${\cal C}$. Assuming that $y$ does not exceed the value
$\frac{2-\sqrt{2}}{2} \approx 0.3$, we deduce from \eq{impi},
\beq
&-& \frac{1}{\nu} {\rm Im} \Pi = - \Theta (w_2 (z) -w) \Theta (w - w_4 (z))
\int_{-z-y}^{z-y} \!\! d \tau {\cal L} (\tau) \nonumber \\
&-& \Theta (w_1 (z) -w) 
\Theta (w - w_2 (z)) \int_{-z-y}^{\frac12 [1-\sqrt{1+4 w}]} \!\! 
d \tau {\cal L} (\tau)   \\
&+& \Theta (w_3 (z) -w) 
\Theta (w - w_4 (z))\int_{-z-y}^{\frac12 [\sqrt{1-4 w}-1]} \!\!\! 
d \tau {\cal L} (\tau) .\nonumber
\eeq
Applying the table integral \eq{integ23} for $x_1=\tau_1$, 
$x_a = x_2 = \tau_3$, $x_b = x_3 = \tau_4$,
and $x_4 = \tau_2$, we obtain the analytic expression
\beq
\!\!\!&-&\!\!\frac{1}{\nu} {\rm Im} \Pi = - \Theta (w_2 (z) -w) 
\Theta (w - w_4 (z))A(z-y)\nonumber \\
\!\!\!&-&\!\!\Theta (w_1 (z) -w) 
\Theta (w - w_2 (z))A(\frac12 [1-\sqrt{1+4w}])\label{impifin} \\
\!\!\!&+&\!\! \Theta (w_3 (z) -w) \Theta (w - w_4 (z))A(\frac12
[\sqrt{1-4w}-1]), \nonumber 
\eeq
where
\beq
w_1 (z) &=& (z+y)^2 + (z+y), \label{omega1} \\
w_2 (z) &=& (z-y)^2 - (z-y), \label{omega2}\\
w_3 (z) &=& -(z+y)^2 + (z+y), \label{omega3}\\
w_4 (z) &=& -(z-y)^2 - (z-y),\label{omega4}
\eeq
and
\beq
A (x) &=& \frac{1}{2 z} \sqrt{\frac{w-z x}{w+ z x} [z^2 - (x+y)^2]} 
+ \frac{k}{4 z \sqrt{w}}
\times \nonumber \\
&\times & [ ((w/z -y)^2 -z^2) F (\varphi (x),k)  \nonumber \\
& & - ((w/z +z)^2 - y^2)  E (\varphi (x),k) \nonumber \\
& & + 2 y (w/z - y-z) \Pi (\varphi (x),n, k)].
\eeq
The argument $\varphi (x)$ and the parameters $k$ and $n$ of the elliptic 
functions $F$, $E$ and $\Pi$ are defined in the following way:
\beq
\varphi (x) &=& \arcsin \sqrt{\frac{(x+z+y) (w/z-y +z) }{2 (z x+w)}} , \\
k &=& \frac{2 \sqrt{w}}{\sqrt{(w/z+z)^2 -y^2}}, \quad n = \frac{2
z}{w/z+z -y}.
\eeq
Note that $\varphi (-z-y) = 0$ and $\varphi (z-y) = \frac{\pi}{2}$, and hence
$A (z-y)$ is expressed in terms of the complete elliptic integrals 
$F (\frac{\pi}{2},k)$, $E (\frac{\pi}{2},k)$, and $\Pi (\frac{\pi}{2},n,k)$.

\subsection{Comparison with the approximate result}

It is also worthwhile to compare our exact result \eq{impifin} with an
approximation for small $z$ commonly used in the literature
(cf., e.g., Refs.~\onlinecite{MCE,MH,SL}). It neglects the square
of the transferred momentum $q^2$ in the denominator of \eq{basicdef2}, and
therefore leads to the kinematic extension of the conventional particle-hole
continuum to a strip $y-y^2 < w <y+y^2$ parallel to the $z$ axis. This
approximation can be effectively expressed in the form
\be
\Pi (q, \omega) \approx - i \frac{q^2}{e^2 \omega} \sigma (0,\omega),
\label{magapp}
\ee
and considered as stemming from the identity
\be
\Pi (q, \omega) = - i \frac{q^2}{e^2 \omega} \sigma (q,\omega),
\ee
with $\sigma (q,\omega)$ replaced by $\sigma (0,\omega)$.
The optical conductivity $\sigma (0,\omega) \equiv \sigma_{\omega}$ can be
easily found \cite{MCE}. For example, its real part equals
\be
{\rm Re} \sigma_{\omega} = \frac{e^2}{16 \pi}\Theta (y^2 - |w-y|).
\label{optcond}
\ee
Thus, Eq. \eq{magapp} yields
\be
-\frac{1}{\nu} {\rm Im} \Pi \approx \frac{4 \pi z^2}{e^2 w} 
{\rm Re} \sigma_{\omega} = \frac{\pi z^2}{4 w} \Theta (y^2 - |w-y|).
\label{magfin}
\ee

We would like to argue that the approximation \eq{magfin} works well
only in a quite small region restricted by the conditions $w_3 (z) <
w <w_2 (z)$ (see Fig.~\ref{optrange}). Since the parabolas $w_2 (z)$
and $w_3 (z)$ intersect at $z = \frac12 [1-\sqrt{1-4 y^2}] \approx
y^2$, this region is represented by a small triangle between the points
$(0,y-y^2)$, $(0,y+y^2)$, $(y^2,y)$. Inside this triangle $-{\rm Im}
\Pi/\nu$ is entirely determined by $- A (z-y)$. Observing that $k^2
\approx 2 n \approx 4 z^2/w$ for small $z<y^2$ and expanding the
complete elliptic integrals with respect to $z$, we find an
asymptotic value,
\be
- A (z-y) \approx  \frac{\pi z^2}{4 w},
\ee
which coincides with \eq{magfin}, except for the domain of applicability.

\begin{figure}[t]
\includegraphics[width=8.6cm,angle=0]{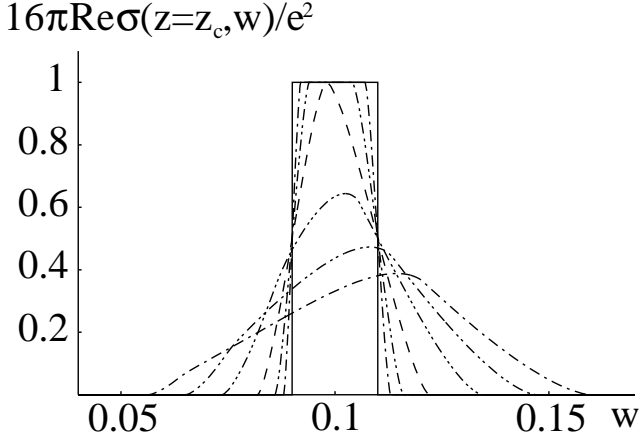}
\caption{The real part of conductivity for the values of
$z_c=0.0$, $0.025$, $0.005$, $0.01$, $0.02$, $0.03$, $0.04$ 
calculated at $y=0.1$.
The unit step corresponds to $z_c=0$. The dashed line corresponds 
to $z_c=y^2=0.01$, which is the limiting value for the applicability 
of the approximation $\sigma (z,w) \approx \sigma (0,w)$.}
\label{optcondsect}
\end{figure}

In Fig.~\ref{optcondsect} we compare the curves 
$16 \pi {\rm Re} [\sigma (z_c,w)]/e^2$
calculated at the different values of $z_c$ and $y=0.1$, 
and plotted as a function of $w$.
The unit-step between $y-y^2=0.09$ and $y+y^2=1.01$ corresponds 
to the optical conductivity \eq{optcond}, which is recovered at $z_c=0$.
One can observe that for $z \ll y^2$ the approximation 
$\sigma (z,w) \approx \sigma(0,w)$
leading to \eq{magfin} works quite well, while for $z \gg y^2$ 
it completely breaks down.

\section{Static limit of the polarization operator}
\label{static}
\subsection{Careful derivation of the static limit}
In this section we derive the static limit
\be
\lim_{w \to 0} {\rm Re} \Pi (z,w) = \lim_{w \to 0} {\rm Re} \Pi^I (z,w)
+ \lim_{w \to 0} {\rm Re} \Pi^{II} (z,w).
\ee
After Eq.~\eq{thesumpi} we have already made an observation that
$\lim_{w \to 0} {\rm Re} \Pi^{II} (z,w)$ might seem to vanish. Below we prove
that, in fact, it does not vanish, but rather produces a finite and a very
important contribution to $\lim_{w \to 0} {\rm Re} \Pi (z,w)$ for small $z<y$.

Let us first identify the quantity $\lim_{w \to 0} {\rm Re} \Pi^I (z,w)$.
One can see that if we put $w=0$ in \eq{roots}, all the roots
$\lambda_{i\sigma}$ become real for all values of the integration variable
$v$. Tracing back the derivation of the expression \eq{funf}, it is easy to
see that $\lim_{w \to 0} {\rm Re} \Pi^I (z,w)$ is obtained, in fact, from the
definition (cf. Ref.~\onlinecite{Raikh})
\be
\widetilde{\Pi}_{q, \omega+ i \delta =0} = \sum_{\chi, \mu =\pm}
\int \frac{d^2 {\bf k}}{(2 \pi)^2}\,\,
\frac{n_F (\epsilon_{{\bf k}}^{\mu}) -
n_F (\epsilon_{{\bf k}+{\bf q}}^{\chi \mu})}
{\epsilon_{{\bf k}}^{\mu} - \epsilon_{{\bf k}+{\bf q}}^{\chi \mu}} 
{\cal F}^{\chi}_{{\bf k}, {\bf k}+{\bf q}}.
\label{raikhdef}
\ee

We would like to emphasize that, in general, 
$\widetilde{\Pi}_{q, \omega+ i \delta =0}$
is not always the same as $\lim_{\omega \to 0} \Pi_{q \omega}$, 
and there might occur a specific situation when
\be
\lim_{\omega \to 0} \Pi_{q \omega} \neq 
\widetilde{\Pi}_{q, \omega+ i \delta =0} .
\label{noneq}
\ee
Once $\lim_{w \to 0} {\rm Re} \Pi^{II} (z,w)$ is nonzero, 
the subtle property \eq{noneq},
for example, holds for the 2DEG with Rashba SO coupling described 
by the Hamiltonian \eq{initham}.

The correct static limit is, of course, given by 
$\lim_{\omega \to 0} \Pi_{q \omega}$.
However, it is very tempting to put $\omega =0$ from the very beginning.
In principle, one can do that. But then, in order to protect oneself 
from a possible mistake,
it is necessary to keep small $\delta \neq 0$ until the very end, even 
during a calculation of a
real part of the polarization operator. More formally, the sequences 
of operations,
\beq
\lim_{\omega \to 0} \int d^2 {\bf k} \lim_{\delta \to 0}
\left ( \cdots \right) \quad {\rm and} \quad
\lim_{\delta \to 0} \int d^2 {\bf k} \lim_{\omega \to 0}
\left ( \cdots \right) \label{r1}
\eeq
always lead to a correct static limit, while the sequence
\be
\int d^2 {\bf k} \lim_{\omega, \delta \to 0} \left ( \cdots \right) 
\label{w1}
\ee
might give in some specific cases an unphysical result breaking causality
and violating analytic properties of a (retarded) response function.

Let us rigorously prove the statements which have been made previously.
In studying the limit $w \to 0$ of 
${\rm Re} \Pi = {\rm Re} \Pi^I + {\rm Re} \Pi^{II}$
it is sufficient to consider how we approach the axis $w=0$ 
from the domains ${\cal D}$
and ${\cal A}$ (see Fig.~\ref{domains}) that cover the ranges 
of momenta $z \in (0,y)$
and $z \in (y,+\infty)$, respectively, as $w \to 0$. In turn, 
the domains ${\cal C}$
and ${\cal B}$ in this limit shrink to the points $z=0$ and $z=y$, 
and therefore their
consideration is not important for our current purpose.

Let us first consider ${\rm Re} \Pi^I$ in the domain ${\cal D}$ at 
very small $w$.
In the range of integration $[\tau_2, \tau_1]$ we can replace
\be
{\cal R}^{\pm} (\tau) \approx {\cal R}_1 (\tau) \,\, 
{\rm sign} (\tau \mp w y/z^2)= \mp {\cal R}_1 (\tau),
\ee
where
\be
 {\cal R}_1 (\tau) = \frac{1}{2 z} 
\frac{\sqrt{y^2 -z^2}}{\sqrt{(w/z)^2 -\tau^2}}.
\ee
Meanwhile, in  the range of integration $[\tau_4, \tau_3]$ we have
\be
{\cal R}^{\pm} (\tau) \approx {\cal R}_2 (\tau) = -\frac{1}{2 z \tau} 
\sqrt{z^2 - (\tau+y)^2}.
\ee
We observe that $\tau_{++} (-)= -\tau_{-+} (+) \approx 1$,
$\tau_{++} (+) = - \tau_{-+} (-) \approx w$, $\tau_{++} (y) \approx w/y$,
$\tau_{-+} (y) \approx -y$; $\tau_{+-} (-)= -\tau_{--} (+) \approx 1$,
$\tau_{--} (-) = - \tau_{+-} (+) \approx w$, $\tau_{+-} (y) \approx - w/y$, 
$\tau_{--} (y) \approx -y$;
$\tau_{++} (0)= -\tau_{-+} (0) = \sqrt{w}$.
Taking into account that $z<y$, we  carefully arrange the limits 
of integration, and thus obtain
\beq
&-& \frac{1}{\nu} \lim_{w \to 0} {\rm Re} \Pi^I =  \nonumber \\
&=& \left( \int_{-y}^{z-y} + \int_{-y}^{-z-y} + \int_{z-y}^{-z-y} 
+ 2 \int_{-y}^{z-y} \right) d \tau {\cal R}_2 (\tau) \nonumber \\
&-& \lim_{w \to 0} \left( \int_{w/y}^{w/z} + \int_{w/y}^{w} + 
\int_{-w/z}^{-w} \right) d \tau  {\cal R}_1 (\tau) \nonumber \\
&+& \lim_{w \to 0} \left(\int_{w/z}^{w} + \int_{-w/z}^{-w} + 
2 \int_{-w/y}^{-w/z} \right)  d \tau {\cal R}_1 (\tau) \nonumber \\
&=& 2 \left( \int_{-y}^{z-y} \!\!+ \int_{-y}^{-z-y}\right) 
d \tau {\cal R}_2 (\tau) - \lim_{w \to 0} 4 \int_{w/y}^{w/z} \!\! 
d \tau {\cal R}_1 (\tau)  
\nonumber  \\
&=& -\int_0^z \frac{2 \tau \sqrt{z^2 -\tau^2} d \tau}{z (\tau^2 -y^2)} - 
\int_{z/y}^1 \frac{2 \sqrt{y^2-z^2} d \tau}{z \sqrt{1 -\tau^2}} \nonumber \\
&=& 2 - \frac{\pi}{z} \sqrt{y^2 -z^2},
\label{repst1}
\eeq
by virtue of the table integral \eq{it1}.

Let us now consider ${\rm Re} \Pi^I$ in the domain ${\cal A}$ at very
small $w$. Taking into account that now $z>y$, we obtain
\beq
&-& \frac{1}{\nu} \lim_{w \to 0} {\rm Re} \Pi^I =  \lim_{w \to 0} 
\left(\int_{w/y}^{\min (1,z-y)} + \int_{w/y}^{\max (w,w/z)} \right. 
\nonumber \\
&-& \int_{-y}^{-\max (w,w/z)} - \int_{-y}^{-\min (1,z+y)} 
+ \int_{\sqrt{w}}^{\min (1,z-y)} \nonumber \\
&+& \int_{\sqrt{w}}^{\max (w, w/z)} - \int_{-\sqrt{w}}^{-\max (w, w/z)} 
- \int_{-\sqrt{w}}^{-\min (1, z+y)}\nonumber \\
&-& \left.  2 \int_{-w/y}^{-\sqrt{w}} - 2 \int_{-y}^{-\sqrt{w}}\right) 
d \tau {\cal R} (\tau) \nonumber \\
&=&  \lim_{w \to 0} \left( \int_{w/y}^{\max (w,w/z)} - 
\int_{-w/y}^{-\max (w,w/z)} - \int_{-y}^{-\min (1,z+y)} \right. \nonumber \\
&+& \left. \int_{\sqrt{w}}^{\min (1,z-y)} - \int_{-y}^{-\sqrt{w}}\right) 
d \tau 2 {\cal R} (\tau).
\label{5int}
\eeq
In the first and the second integrals of the last part of \eq{5int}, 
we can replace
\be
{\cal R} (\tau) \approx {\cal \tilde{R}}_1 (\tau) = - \frac{1}{2 z} 
\frac{\sqrt{z^2 -y^2}}{\sqrt{\tau^2 - (w/z)^2}},
\ee
while in the third integral
\be
{\cal R} (\tau) \approx {\cal R}_2 (\tau) \,\, {\rm sign (\tau)}.
\label{zamsp}
\ee
The fourth and fifth integrals require a more careful consideration.
We note that for small $w$ there holds the inequality $\sqrt{w} > w/z$,
which allows us to neglect $w$ in ${\cal R} (\tau)$. Therefore these two 
integrals can be rewritten
as
\beq
& & 2 \lim_{w \to 0} \left( \int_{-y}^{\min (1,z-y)}  - 
\int_{-\sqrt{w}}^{\sqrt{w}}\right) d \tau {\cal R}_2 (\tau) \nonumber \\
& & = 2  \int_{-y}^{\min (1,z-y)} d \tau {\cal R}_2 (\tau),
\label{osobint}
\eeq
where the integral on the rhs of \eq{osobint} is understood in the 
sense of principal value.

Thus we obtain for $z>y$
\beq
&-& \frac{1}{\nu} \lim_{w \to 0} {\rm Re} \Pi^I =-\int_{z/y}^{\max (1,z)}
\frac{2 \sqrt{z^2 -y^2} d \tau}{z \sqrt{\tau^2 -1}} \label{repst2} \\
&-& \!\! \int_{0}^{\min (1-y,z)}\frac{\sqrt{z^2 -\tau^2} d \tau}{z (\tau+y)} 
- \int_{0}^{\min (1+y,z)}\frac{\sqrt{z^2 -\tau^2} d \tau}{z (\tau-y)} . 
\nonumber
\eeq
It is easy to calculate this expression using the table integral  \eq{it2}.
The result is expressed in \eq{it21} and \eq{it22}.

Let us now consider the static limit of ${\rm Re} \Pi^{II}$.
In Appendix \ref{statapp} we show that it equals
\be
- \frac{1}{\nu} \lim_{w\to 0} {\rm Re} \Pi^{II}=
\frac{\pi \sqrt{y^2-z^2}}{z}\Theta (y-z).
\label{correct}
\ee
We note that this term is nonzero only for $z<y$ (or $q< 2 k_R$).
One can observe that \eq{correct} is exactly canceled by the
counterterm from \eq{repst1}.

Collecting all contributions and introducing $\sin \psi = y/z$ for $y<z$
and $\sin \psi_{\pm} = (1 \pm y)/z$ for $1 \pm y <z$, we present the 
static limit of the polarization operator in the form
\begin{widetext}
\beq
&-& \frac{1}{\nu} {\rm Re} \Pi (z,0) = 2  \,\, \Theta (1-y-z) +
\Theta (y-|z-1|) \,\, (1 +\frac{\pi}{2} \sin \psi) - 2 \, \Theta (z-1) \,
{\rm arccosh} z \, \cos \psi \nonumber \\
&+& \sum_{\mu = \pm} \Theta (z-(1 + \mu y)) \left(  1+ \mu \psi_{\mu} 
\sin \psi - \cos \psi_{\mu} -
2 \cos \psi \ln \frac{1 + z \sin (\psi_{\mu} - \mu \psi)}{2 \sqrt{2 z} 
\cos \frac12 \psi_{\mu} \cos \frac12 \psi}\right),
\label{eq5}
\eeq
\end{widetext}
where ${\rm Re} \Pi (z,0) \equiv \lim_{w \to 0} {\rm Re} \Pi (z,w)$.
In what follows we also use the abbreviations  ${\rm Re} \Pi^I (z,0)$ and
${\rm Re} \Pi^{II} (z,0)$ for $\lim_{w \to 0} {\rm Re} \Pi^I (z,w)$ and 
$\lim_{w \to 0} {\rm Re} \Pi^{II} (z,w)$.

\subsection{Analysis of Eq.~\eq{eq5}}
 Let us analyze the expression \eq{eq5}. For the values
$z<1-y$ we obtain ${\rm Re} \Pi (z,0) = - 2 \nu$, which ensures the
fulfillment of the compressibility sum rule. For $z>1-y$, ${\rm Re}
\Pi (z,0)$ deviates from the value $-2 \nu$.

\begin{figure}[b]
\includegraphics[width=8.4cm,angle=0]{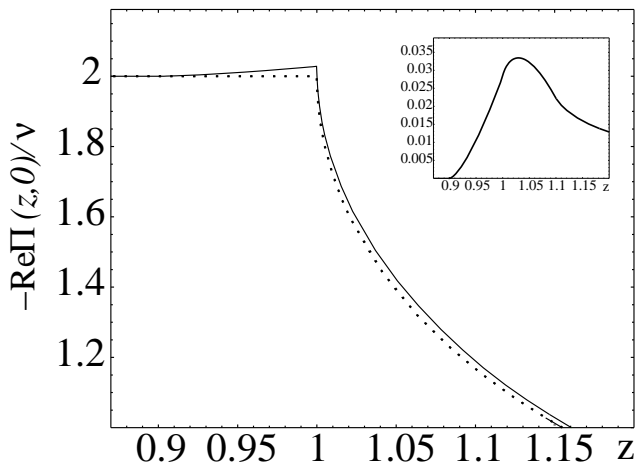}
\caption{The behavior of $-{\rm Re} \Pi (z,0)/\nu$ near  $z=1$ for $y=0.1$ 
(solid line)
and $y=0$ (dotted line). The difference between the two curves is shown 
in the inset.}
\label{2kfplot}
\end{figure}

For $y=0$, Eq.~\eq{eq5} reproduces the conventional result of 
Stern, \cite{Stern}
\be
-\frac{1}{\nu} {\rm Re} \Pi_{y=0} (z,0) = 2 -2 \sqrt{1-1/z^2} \Theta (z-1).
\ee
In Fig.~\ref{2kfplot} we compare $-{\rm Re} \Pi (z,0)/\nu$ for $y=0.1$ 
and $y=0$.
We demonstrate that $-{\rm Re} \Pi_{y \neq 0} (z,0)/\nu$ is always larger than
$-{\rm Re} \Pi_{y = 0} (z,0)/\nu$, although the difference between the 
two curves
(shown in the inset) is quite small. In particular, the maximal value of
$-{\rm Re} \Pi (z,0)/\nu$ at $z=1$ scales with $y$ like
\be
-\frac{1}{\nu} {\rm Re} \Pi (1,0) \approx 2+ \frac{2 \sqrt{2}}{3} y^{3/2},
\label{scz1}
\ee
while at large $z$ the asymptotic behavior of \eq{eq5} is $\sim
\frac{1+y^2}{z^2}$.

An important test of our results is provided by  the Kramers-Kronig relations
and the sum rules. For example, using the analytic expressions for ${\rm Re}
\Pi (z,0)$  and ${\rm Im} \Pi (z,w)$ we have checked the
zero-frequency Kramers-Kronig relation
\beq
{\rm Re} \Pi (z, 0 ) &=& \frac{2}{\pi} \int_0^{\infty} \frac{d w}{w} \,\,
{\rm Im} \, \Pi (z, w). \label{compress}
\eeq
The actual limits of the integration are finite and given by the boundaries
of the particle-hole continuum. This has allowed us to confirm \eq{compress}
numerically with a sufficiently good accuracy (the deviation is $\leq 10^{-6}$
even for a quite simple routine).

\begin{figure}[t]
\includegraphics[width=8.4cm,angle=0]{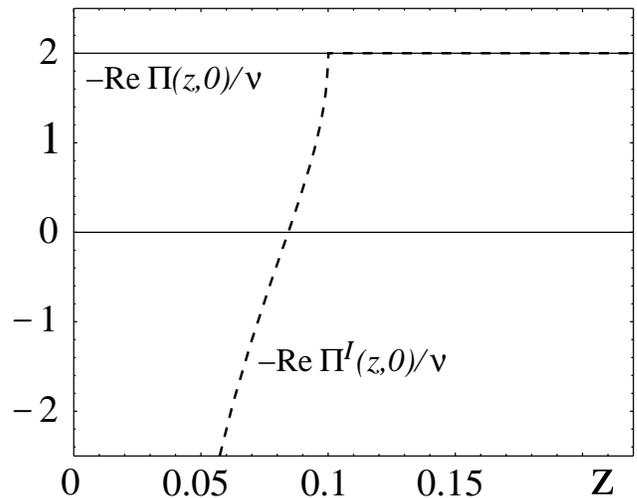}
\caption{The behavior of $-{\rm Re} \Pi (z,0)/\nu$ (solid line) and 
$-{\rm Re} \Pi^I (z,0)/\nu$ (dashed line) near $z=y$ for $y=0.1$.}
\label{2krplot}
\end{figure}

It is instructive to compare ${\rm Re} \Pi (z,0)$ and ${\rm Re} \Pi^I (z,0)$,
\be
-\frac{1}{\nu} {\rm Re} \Pi^I (z,0) =  -\frac{1}{\nu} {\rm Re} \Pi (z,0) -
\frac{\pi \sqrt{y^2-z^2}}{z}\Theta (y-z).
\label{incorr2}
\ee
It is obvious that ${\rm Re} \Pi$ does not have any anomaly at $z=y$
(cf. Ref.~\onlinecite{Raikh}), while ${\rm Re} \Pi^I$ does have it 
(see Fig.~\ref{2krplot}).
One can also observe that ${\rm Re} \Pi^I$ diverges in the limit 
$z \to 0$ and changes
the sign at some value of $z$. Therefore, it cannot be regarded itself 
as a correct static limit.
Otherwise, it would have violated the compressibility sum rule and generated 
an instability of the medium by virtue of SO coupling.

Thus, we have demonstrated that in order to obtain the correct static limit of
the polarization operator for the 2DEG with Rashba SO coupling it is crucially
important to follow the thorough definition that implies \eq{r1}, but not
\eq{w1}. Otherwise, the contribution ${\rm Re} \Pi^{II} (z,0)$ is missing.
Being the difference between ${\rm Re} \Pi (z,0)$ and ${\rm Re} \Pi^I (z,0)$,
it shows up for $z<y$ (or $q<2 k_R$), and disappears only when $k_R$ goes to
zero. Therefore, for the conventional 2DEG ($k_R =0$) there is no difference
between $\Pi_{q,0}$ and $\widetilde{\Pi}_{q,0}$ [and between \eq{r1} and
\eq{w1} as well]. So the property \eq{noneq} is a peculiar feature of the 2DEG
with $k_R \neq 0$. We suppose that it might be related to the singularity of
the spectrum \eq{SOspectrum} at $k=0$.

\subsection{Effective interaction: modification of the Friedel
$2k_{F}$-oscillations}
 Let us conveniently rewrite the RPA expression
\eq{rpa} for the dielectric function in terms of $\tilde{r}_s =
\frac{r_s}{2\sqrt{2}}$,
\be
\varepsilon (z,w) = 1-\frac{\tilde{r}_s}{z \nu} \Pi (z,w),
\label{rpa1}
\ee
where $r_s = \frac{\sqrt{2} m^* e^2}{k_F \varepsilon_{\infty}}$ is the 2D
Wigner-Seitz parameter. The latter controls the accuracy of RPA, which 
becomes better with decreasing $r_s$.
In the limit $w \to 0$, Eq.~\eq{rpa1} describes the static screening 
of the Coulomb interaction.
It is well-known \cite{Stern} that the singular behavior of the derivative,
\be
-\frac{1}{\nu} \frac{d}{dz} \Pi_{y=0} (z,0) \big|_{z=1+\alpha}
\approx -\sqrt{\frac{2}{\alpha}},
\ee
at $z = 1+\alpha$ with small $\alpha>0$ gives rise to the Friedel 
oscillations of a screening potential, their leading asymptotic term being
\be
V_{osc}^{y=0} (\rho) = - \frac{2 e^2 k_F}{\varepsilon_{\infty}} \cdot 
\frac{2 \tilde{r}_s e^{-2 k_F d}}
{[\varepsilon_{y=0} (1,0)]^2} \frac{\sin 2 k_F \rho}{(2 k_F \rho)^2}, 
\label{vosc}
\ee
where $\rho = \sqrt{x_1^2 +x_2^2}$, and $d$ is a distance from a  
probe charge
to the 2DEG plane. The SO modification to \eq{vosc} follows from
\be
-\frac{1}{\nu}  \frac{d}{dz} \Pi (z,0) \approx - \sqrt{\frac{2}{\alpha}} 
\cdot \sqrt{1-y^2} ,
\ee
and results in
\be
V_{osc} (\rho) = - \frac{2 e^2 k_F}{\varepsilon_{\infty}} \cdot 
\frac{2 \tilde{r}_s e^{-2 k_F d}}
{[\varepsilon_{y=0} (1,0)]^2} \frac{\sin 2 k_F \rho}{(2 k_F \rho)^2} Q (y),
\label{voscy}
\ee
where the factor
\be
Q (y) = \sqrt{1-y^2} 
\left[ \frac{\varepsilon_{y=0} (1,0)}{\varepsilon (1,0)}\right]^2
\ee
is always smaller than $1$. It means that due to the SO coupling the 
amplitude of the Friedel
oscillations is diminished (cf. Ref.~\onlinecite{Raikh}), although the amount 
of such decrease is
quite small ($\sim 0.5 \%$ for $y=0.07$ and $r_s =0.2$). We note that due 
to \eq{scz1} it is possible to approximate
\be
\left[ \frac{\varepsilon_{y=0} (1,0)}{\varepsilon (1,0)}\right]^2  
\approx 1-\frac{4 \sqrt{2} y^{3/2}}{3 (1 +2 \tilde{r}_s)}.
\ee
Rigourously speaking, it is necessary to take into account in \eq{voscy} 
the dependence of $k_F$ on $k_R$ as well.

The second derivative $\frac{d^2}{dz^2} \Pi (z,0)$ also diverges at the 
points $z=1 \pm y+0^+$,
thus contributing to the subleading asymptotic terms of the oscillating 
potential~\cite{Raikh}.

\section{Structure factor and SO-induced damping of plasmons}
\label{structure}

An important quantity which either can be directly observed or
enters into expressions for other observable quantities is the
structure factor~\cite{Mahan}
\be
S (z, w)=- {\rm Im} \left[1/\varepsilon(z,w) \right].
\ee
It depends on the Wigner-Seitz parameter $r_s$ and contains
information about both particle-hole excitations and collective
excitations (plasmons). The spectrum of the latter is found from
the equation ${\rm Re} \, \varepsilon =0$.
For the conventional 2DEG it can be derived on the basis of 
Ref.~\onlinecite{Stern} 
and equals\cite{Singwi}
\beq\label{plas}
w_{pl}(z)=\frac {z(z+2\tilde{r}_{s})}{2\tilde{r}_{s}}
\sqrt{\frac{4\tilde{r}_{s}^{2}+
4\tilde{r}_{s}z^{3}+z^{4}}{z(z+4\tilde{r}_{s})}} \Theta (z^* -z),
\eeq
where the endpoint of the spectrum $z^*$ is the real positive root of
the equation $z^2 (z+ 4\tilde{r}_{s}) = 4 \tilde{r}_{s}^2$.
Provided ${\rm Im} \, \varepsilon =0$, the plasmon spectrum is undamped, 
and the structure factor
is $\delta$-peaked at $w = w_{pl}(z)$,
\be
S_{pl} (z,w) = \alpha (z) \delta (w - w_{pl}(z)),
\ee
with the weight factor  $\alpha (z) = \pi [{\rm Re}\, 
\partial\varepsilon/\partial w|_{w=w_{pl} (z)}]^{-1}$, given by
\beq\label{plasmon}
\alpha(z)  = \frac{\pi\sqrt{z}[16\tilde{r}_{s}^{4}-z^{4}
(z+4\tilde{r}_{s})^{2}]}{2 \tilde{r}_{s}
\sqrt{(4\tilde{r}_{s}^{2}+4\tilde{r}_{s}z^{3}+z^{4})(z+4\tilde{r}_{s})^{3}}}.
\eeq
The plasmon spectrum can be visualized on a contour plot of $S (z,w)$ by 
adding an artificial
infinitesimal damping $\delta$ to ${\rm Im} \, \varepsilon$.
It becomes very helpful when an exact analytic expression similar to 
\eq{plas} is not known.

In the presence of SO coupling, the structure factor depends on $y$ as well.
Solving approximately the equation ${\rm Re} \, \varepsilon =0$ at $y\neq 0$, 
we can establish how the plasmon spectrum is modified by SO coupling. In fact, 
we observe that the SO modification of \eq{plas} is quite small. Later on, we 
will comment on how small it is, and explain why this effect is not really 
important (cf. Ref.~\onlinecite{wang}).

A much more important effect is a damping of plasmons generated by SO coupling.
As it has been already discussed, SO coupling extends the continuum of the 
particle-hole
excitations up to the new boundaries shown in Fig.~\ref{optrange}. Therefore, 
the plasmon spectrum is expected to acquire a finite
width, whenever it enters into the SO-induced region of damping.

\begin{figure}[t]
\includegraphics[width =4.25cm]{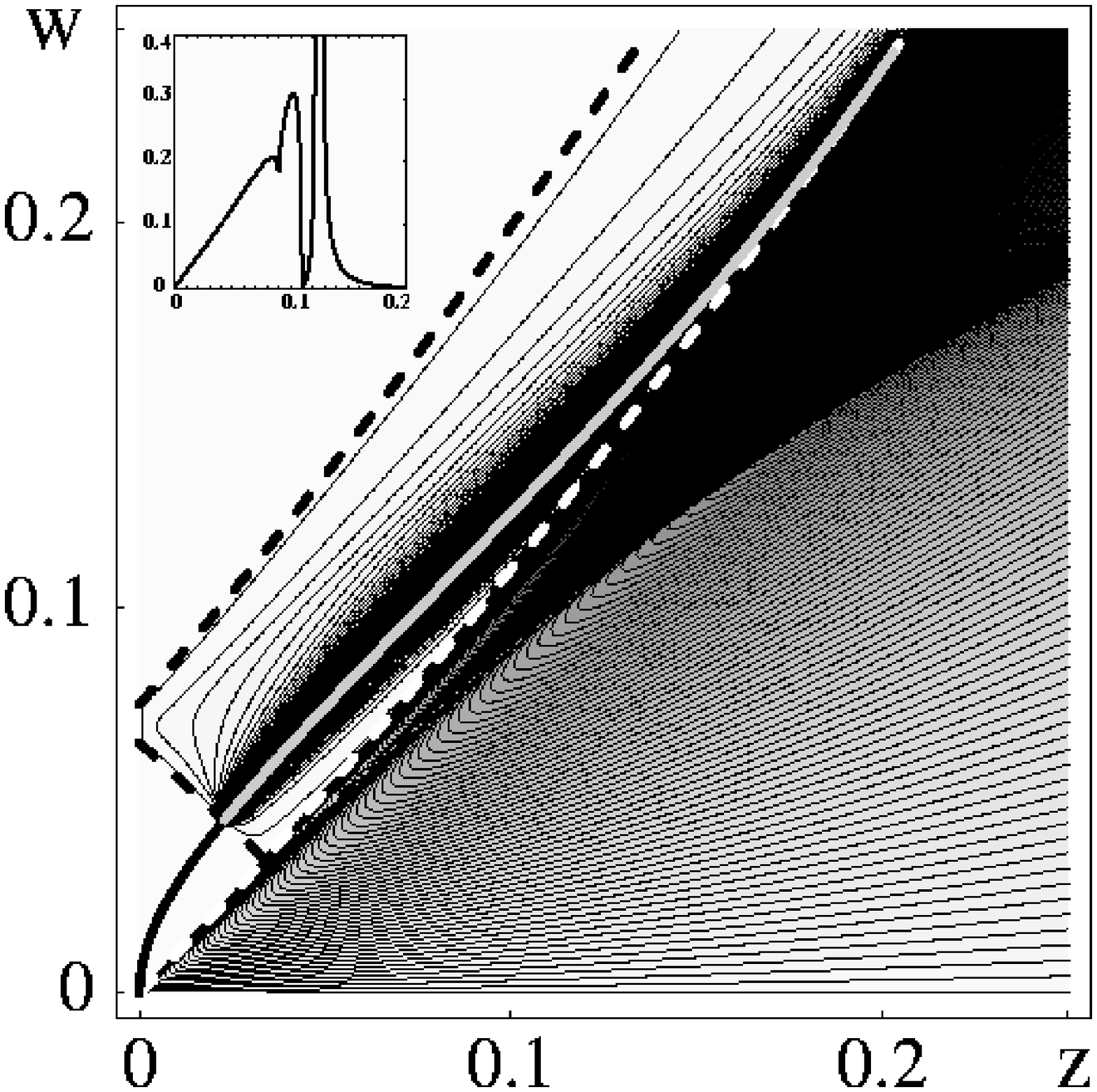}
\includegraphics[width =4.25cm]{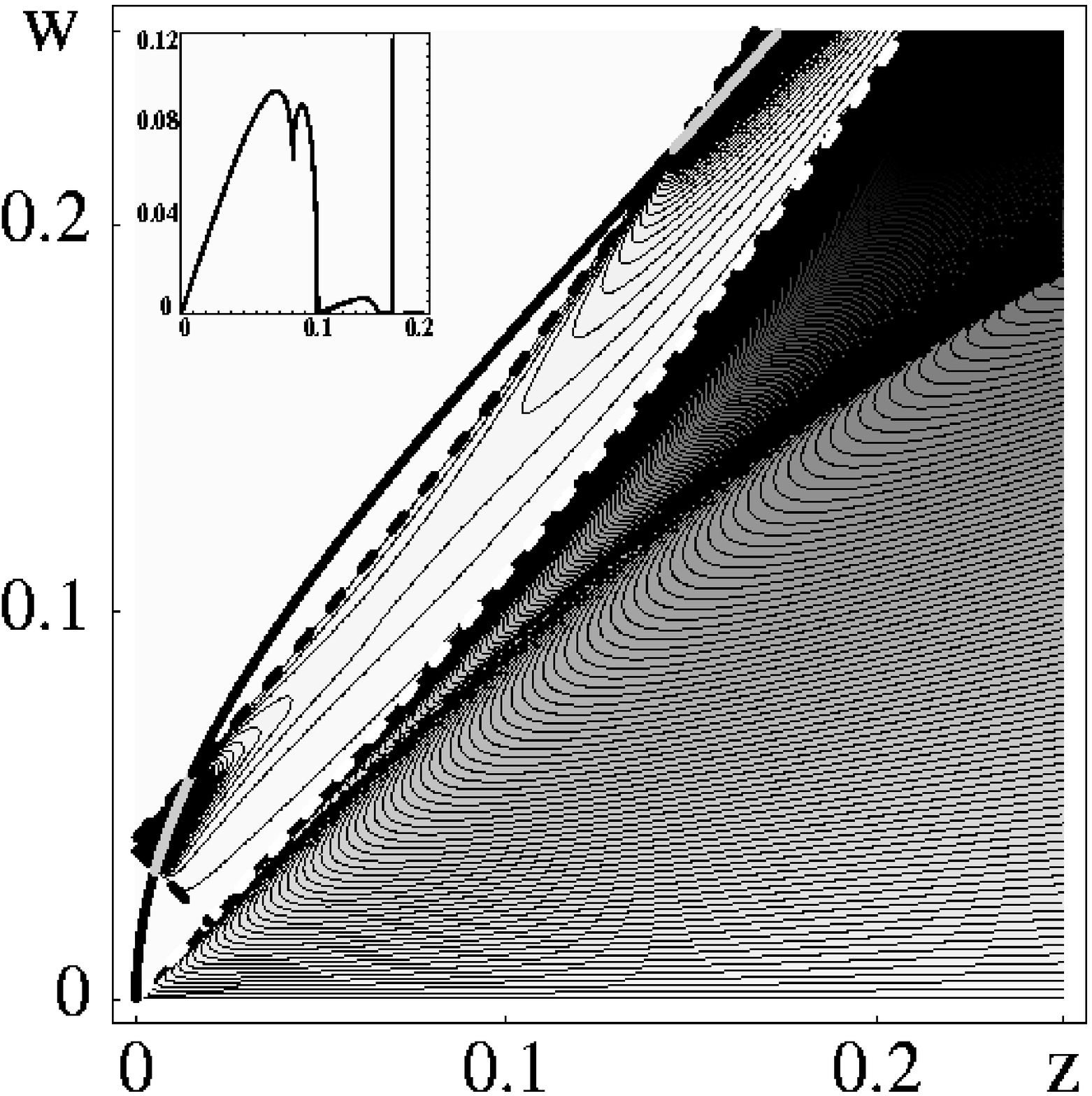}
\caption{Contour plots of $S (z,w)$ showing
the SO-induced wedge-shaped damping region (bounded by the dashed lines).
The plasmon mode is depicted by the bold line.
Insets show the cross-sections $S(z=0.1,w)$ as a function of $w$.
{\it Left panel:} $y\!=\!0.07$, $r_{s}\!=\!0.2$.
{\it Right panel:} $y\!=\!0.04$, $r_{s}\!=\!0.6$. }
\label{g0}
\end{figure}

In Fig.~\ref{g0} we show the contour-plots of the structure factor $S (z,w)$
depicting the plasmon spectrum  by the bold line, in black, where
${\rm Im} \, \varepsilon  = 0$ (undamped plasmon, the structure factor 
is $\delta$-peaked)
and in gray, where ${\rm Im} \, \varepsilon \neq 0$ (SO-damped plasmon, 
the structure
factor has a finite height and width). Depending on the values of $r_s$ 
and $y$, the two different cases are possible:
(I) the plasmon enters only once into the SO-induced damping region
(left panel);
(II) it enters twice, escaping for a while after the first entrance
(right panel).

Within the conventional boundaries $w=z^2 \pm z$ of the particle-hole
continuum, the structure factor $S (z,w)$ is modified by SO coupling only
slightly and can be approximated by the conventional expressions\cite{Stern}.

\begin{figure}[t]
\includegraphics[width=8.4cm,angle=0]{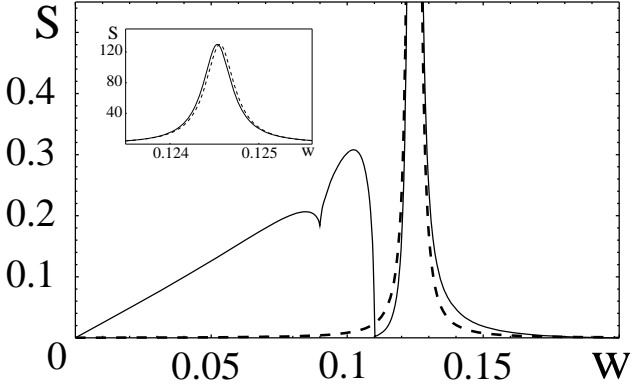}
\caption{The cross section $S(z=0.1,w)$ (solid line) compared to 
the approximation
$S (z=0.1,w)_{SO-damp \,\, pl}$ (dashed line). Parameters: $y\!=\!0.07$, 
$r_{s}\!=\!0.2$.}
\label{lorentzfig}
\end{figure}

In the SO-induced region of damping, $S (z,w)$ is very
well approximated by the Lorentzian function describing the SO-damped 
plasmon with the width $\gamma (z)$,
\beq
S (z,w)_{SO-damp \,\, pl}=\frac{\alpha (z)}{\pi}
\frac{\gamma(z)}{(w-w_{pl}(z))^{2}+\gamma^{2} (z)},
\label{lorentz}
\eeq
where $w_{pl}(z)$ and $\alpha (z)$ are supposed to be practically
independent of $y$ and therefore given by \eq{plas} and \eq{plasmon}, while
\be
\gamma(z)=\frac{\alpha (z)}{\pi} {\rm Im} \, \varepsilon (z , w_{pl} (z))
\ee
essentially depends on $y$ via ${\rm Im} \, \varepsilon \neq 0$.
In Fig.~\ref{lorentzfig} we present the enlarged plot with the cross section
$S (z=0.1,w)$ from the inset of the left panel of Fig.~\ref{g0} and compare
it with the approximate $S (0.1,w)_{SO-damp \,\, pl}$ given by \eq{lorentz}.
The inset of Fig.~\ref{lorentzfig} shows both curves on a more fine scale.
Comparing
the positions of their peaks, we conclude that  the shift of the  plasmon
dispersion due to SO coupling from $w_{pl} (z)$ \eq{plas} is one or two orders
of magnitude smaller than $\gamma (z)$ [unless $\gamma (z)=0$], and therefore
can hardly be resolved experimentally. The almost equal height of the peaks
confirms that \eq{plasmon} is also a very good approximation for the weight
factor at $y \neq 0$.

\begin{figure}[t]
\includegraphics[width =8.65cm,angle=0]{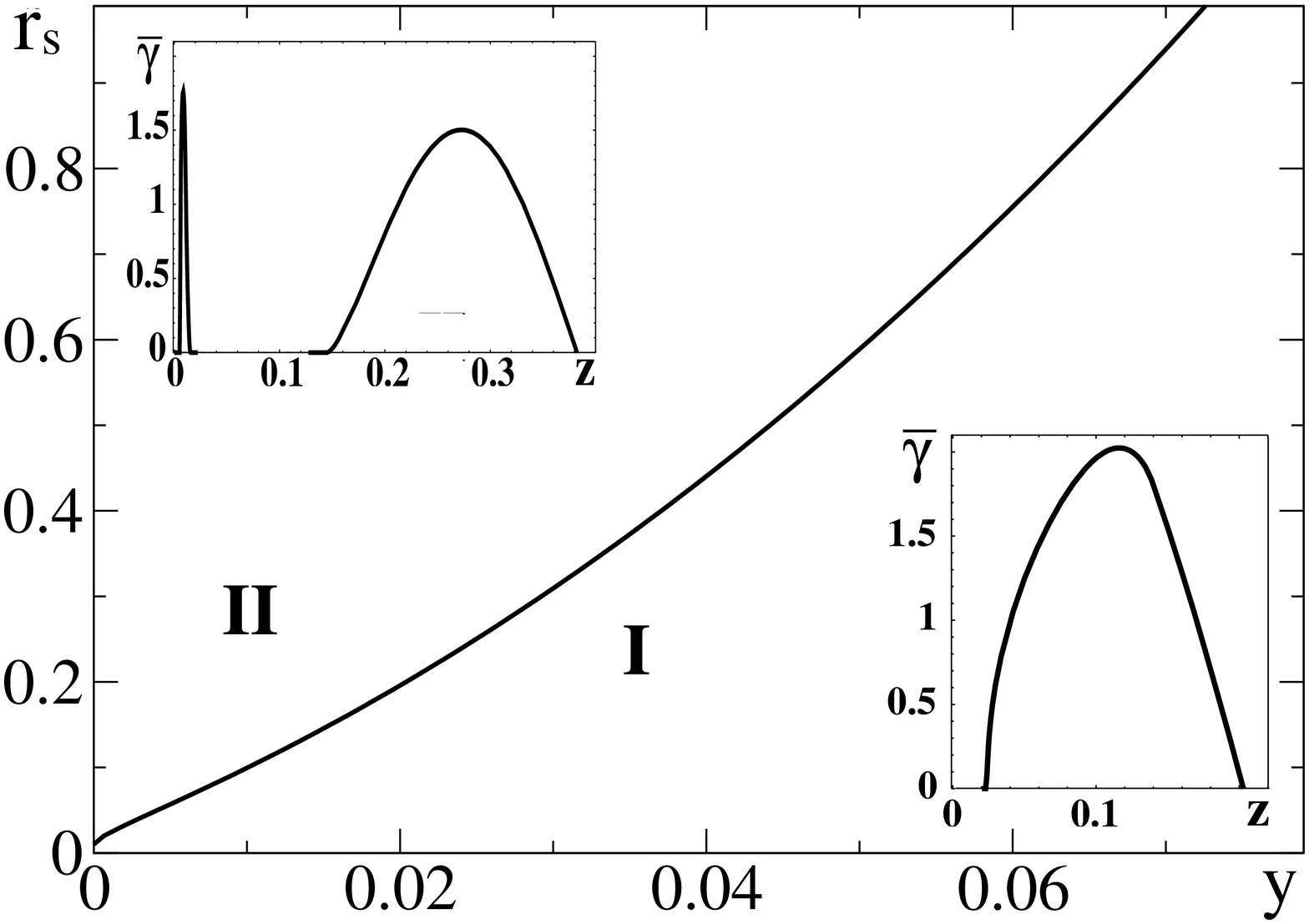}
\caption{The plane $(y,r_{s})$ is divided into the domains I and II, where
the plasmon has one or two undamped pieces, respectively (cf. Fig.~\ref{g0}).
Insets show SO-induced plasmon width
$\overline{\gamma}(z)=\gamma (z)\times 10^4$ for the same parameters
as in Fig.~\ref{g0}.}
\label{crit}
\end{figure}

Let us find the values of $y$ and $r_s$ that correspond to the typical cases
(I) and (II) shown in the left and right panels of Fig.~\ref{g0}, respectively.
For this purpose, it is necessary to establish when the curves $w_1 (z)$
\eq{omega1} and $w_{pl} (z)$ \eq{plas} touch each other.
The result is represented in Fig.~\ref{crit}, where the plane $(y, r_s)$
is divided into the corresponding domains I and II. Changing $y$ and $r_s$,
we can tune the relative position of the SO-induced particle-hole continuum
and the plasmon spectrum. The insets show $\overline{\gamma} (z)=\gamma
(z)\times 10^4$ for the values of parameters the same as in Fig.~\ref{g0}:
(I) $y=0.07$, $r_s=0.2$; (II) $y=0.04$, $r_s=0.6$.
The function $\gamma (z)$ is defined for $z \leq z^*$ and vanishes
where the plasmon is undamped.

A direct measurement of SO-induced plasmon width can be provided by
means of inelastic light (Raman) scattering (see, e.g.,
Ref.~\onlinecite{Raman}). Experimentally it is possible to measure the
structure factor in the range $q=0-2\times 10^{7}$m$^{-1}$ which
corresponds to the range of $z=0-0.1$ (for $r_{s}\sim 1$). Keeping
$z$ at a fixed value and varying $y$ and $r_s$, one should observe
different patterns of $S (z,w)$ with either damped or undamped
plasmon, like those shown in the insets to Fig.~\ref{g0}, where the
sections of $S (z,w)$ at constant $z=0.1$ are presented. In order to
estimate realistic parameters, we focus on InAs-based 2DEG, where the
strength of Rashba coupling $\alpha_{R}$ has quite large values up
to $\sim 3\times 10^{-11}$eVm (Ref.~\onlinecite{large}) and prevails over
the Dresselhaus term \cite{Ganichev}. Most of experiments deal with
the 2DEG's densities ranging from $n=0.7\times 10^{16}$m$^{-2}$
(Ref.~\onlinecite{Engels}) to $n=2.4\times 10^{16}$m$^{-2}$
(Ref.~\onlinecite{Nitta}). Taking $\varepsilon_{\infty} \approx 12$ and
$m^{*} \approx 0.03 m_e$, we obtain the range of $y$ from 0.04 to
0.075 and the range of $r_{s}$ from 0.38 to 0.3. These values belong
to the domain I in Fig.~\ref{crit}, where the SO-damping of the
plasmon is especially important. Choosing the value of $\gamma \sim
10^{-4}$, we establish that $2\tau_{pl}=\hbar/(4\gamma E_{F})$ is of
the order of 10 ps. This can, in principle, be resolved
experimentally, since the Raman measurements done in II-VI quantum
wells \cite{Jus} give a finite lifetime of the plasmon mode with the
typical value $2\tau_{pl} \sim 0.3$ ps. We also point out that the
II-VI compounds (e.g., HgTe) represent an even more promising
candidate for observation of the discussed effects since the
reported SO coupling strength is comparatively large\cite{Mol}.

Plasmon broadening may be also caused by thermal effects.
We use the results for 2DEG at finite temperature \cite{Fetter} in order to
estimate the relative contributions of SO-induced and thermal damping. The
former dominates over the latter below some characteristic temperature that
is about 90 K for the parameters of Ref.~\onlinecite{large}. This estimate is 
made at constant $z=0.1$. Note that the thermal damping strongly depends on 
the values of $r_{s}$ and temperature, whereas the SO-induced damping has quite
a weak $r_{s}$ dependence (e.g., compare the insets in Fig.~\ref{crit}).
This may provide a guide for an experimental separation of these two sources
of broadening.

\begin{figure}[t]
\includegraphics[width=6.0cm,angle=270]{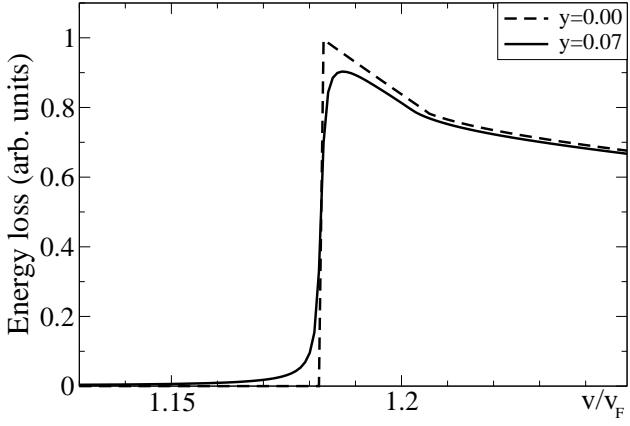}
\caption{The  plasmon contribution to the energy-loss function at $y=0$
(dashed line) compared with the SO-damped plasmon contribution at $y=0.07$
(solid line). For the both curves $r_s=0.2$.}
\label{en_loss}
\end{figure}

A finite lifetime of plasmons would modify the energy-loss
function. The energy loss per unit length for a particle moving toward 
a plane with 2DEG with velocity $v$ is given by
\beq
-\frac{dW}{dx}= C \left( \frac{v_F}{v} \right) \int_0^{\infty} z dz
\int_0^{v/v_F} \frac{ u S (z, u z) du}{\sqrt{(v/v_F)^2 - u^2}},
\label{wloss}
\eeq
where $C=16 \pi^2 e^2 n/\varepsilon_{\infty}$.
In the presence of SO coupling, the most essential modification is expected in
the plasmon sector. In Fig.~\ref{en_loss} we plot the plasmon contributions
$(-dW/dx)_{pl}$ for $y=0$ and $y=0.07$ as a function of $v/v_F$. Both curves
are calculated at $r_s=0.2$ and normalized by the peak value of
$(-dW/dx)_{pl}$ at $y=0$. We
observe that the sharp peak is smoothed down and the steplike
behavior becomes more smeared with increasing SO coupling strength.

\section{SO-induced damping of LO phonons}
\label{phonons}

Our results for the  structure factor allow us to predict that  the other
collective mode, longitudinal optical (LO) phonon, will also experience
a pronounced
SO-induced damping. Its dispersion and lifetime can be obtained from the
renormalized propagator that is expressed in RPA as \cite{JS}
\beq
D(q,\omega)=\frac{2\omega_{LO}}{\omega^{2}-\omega^{2}_{LO}-2\omega_{LO} 
M_{q}^{2}\Pi_{q \omega} /\varepsilon_{q\omega}},
\label{phonon}
\eeq
where $M_{q}^{2}=\textstyle{\frac{1}{2}}  \kappa V_{q} \omega_{LO}$, and
$\kappa=1-\varepsilon_{\infty}/\varepsilon_{0}$ is a material parameter.
On the basis of this
expression, we can  establish that the LO-phonon width
(normalized by $2 k_F^2/m^*$) equals
\be
\gamma_{LO} (z)= \frac{\kappa S (z,w) (w_0 {\rm Re} \,
\varepsilon )^2}{2 w ({\rm Re} \, \varepsilon )^2 + \kappa w_0^2 {\rm Re}
(\partial \varepsilon / \partial w)} \bigg|_{w =w_{ph} (z)},
\ee
where $w_0 = (m^*/2 k_F^2) \,\, \omega_{LO}$ and $w_{ph} (z)$ is the
renormalized phonon spectrum. Like in the case of plasmons, we may
neglect the dependence of $w_{ph} (z)$ on SO coupling, and find the
phonon spectrum from the equation $w^2 = w_0^2 [1 + \kappa (({\rm Re} \, 
\varepsilon )^{-1} -1)]$, where ${\rm Re}
\, \varepsilon$ is taken at $y=0$.

\begin{figure}[t]
\begin{center}
\includegraphics[width=9cm,angle=0]{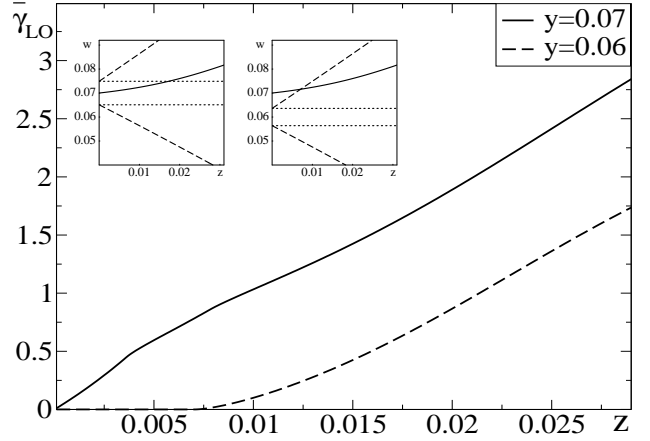}
\caption{
SO-induced damping of the LO phonon
$\overline{\gamma}_{LO} (z) = \gamma_{LO} (z) \times 10^4$ for
$w_0=0.07$, $r_s =0.4$, $\varepsilon_{\infty}=12$, and $\varepsilon_0 =15$.
Solid and dashed lines correspond to the cases $y=0.07$
($|w_0 -y|<y^2$) and $y=0.06$ ($|w_0 -y|>y^2$), respectively.
Insets show the location of the phonon spectrum (solid line)
relative to the SO-induced damping region (dashed lines) and
to the strip $y-y^2 < w<y+y^2$ (dotted lines) for $y=0.07$
(left) and $y=0.06$ (right).}
\label{lifephon}
\end{center}
\end{figure}

The lifetime effects of the LO phonons can be very important for a coupled
dynamics of carriers and phonons.
An interesting possibility arises when $\omega_{LO} \sim 2 \alpha_R k_F$ 
(or $w_0 \sim y$).
This, for example, holds in InAs, where the value
$\omega_{LO} \approx 28$ meV gives $w_0 \approx 0.07$.
Changing the Rashba coupling strength $\alpha_R$ by an applied electric 
field, one
can manipulate the lifetime of an optical phonon, which would result in a 
modification
of transport properties by virtue of the electron-phonon coupling.

In Fig.~\ref{lifephon} we show the function
$\overline{\gamma}_{LO} (z) = \gamma_{LO} (z) \times 10^4$.
In our calculations we used the following parameters:
$w_0 =0.07$, $r_s=0.4$, $\varepsilon_{\infty} = 12$, and $\varepsilon_0 = 15$ 
(for InAs).
Choosing the value $\gamma_{LO} = 10^{-4}$, we find that for LO phonons the
SO-induced lifetime $2 \tau_{LO}$ is of the order of 10 ps as well.
We note that the typical lifetime for the LO phonons  measured in AlAs and GaAs
by the Raman spectroscopy is of the same order \cite{LOphonon}.
One can observe that for $y=0.07$ ($|w_0-y| < y^2$)  the damping is always
finite (solid line), while for $y=0.06$ ($|w_0-y| > y^2$) it becomes nonzero
only after some value $z \approx 0.007$ (dashed line). In the two insets we
show the relative location of the phonon spectrum (solid line) and of the
SO-induced damping region (dashed lines). We also depict the boundaries of the
strip $w = y\pm y^2$
(dotted lines), and one can see that for $y=0.07$ the phonon spectrum leaves it
at the value $z \approx 0.017$ (left inset), while for $y=0.06$ the phonon 
spectrum
never gets inside the strip (right inset). This indicates that the the 
approximation
\eq{magfin} is insufficient for describing the discussed effect.

\section{Conclusion}

We have calculated the dielectric function of 2DEG
with Rashba SO coupling in RPA at zero temperature. We have
described the new features of screening that appear due to SO coupling.
In particular, we have discussed in detail the extension of the region of
the particle-hole excitations, which leads, in turn, to the
additional broadening of the plasmon mode. The same mechanism generates
the SO-induced lifetime of the longitudinal optical phonons.
Speaking generally, we have seen how SO coupling tends to suppress collective
excitations due to the relaxation via intersubband transitions.
At the same time, SO coupling does not affect much the position
of the collective mode dispersions, and therefore this effect is less
important.

Another conclusion that can be derived from our studies is that the 2DEG
with Rashba SO coupling requires a more careful treatment as far as various
approximations are concerned. Usually they fall into the following categories:
(1) the so-called $\xi$-approximation based on the linearization of the
spectrum; (2) an expansion of physical quantities in powers of SO-coupling
strength before their evaluation; (3) reshuffling the
operations of the momentum integration and taking the limit of zero
frequency. The examples that indicate that none of these approximations
provide a fully reliable result for the system in question are the following:
(1) being a version of the $\xi$-approximation, Eq.~\eq{magapp} has a very 
limited range of applicability;
(2) the static polarization operator at small $y$ cannot be obtained by means
of the power expansion in $y$, because its maximal value scales as $y^{3/2}$
[see Eq.~\eq{scz1}]; (3) reshuffling the order of operations leads to the
unphysical result with violated analytic properties. Although the underlying
reason of the failure is not quite clear, the above examples warn against 
the blind usage of these popular approximations in this system.

Finally, the comparison of our theoretical estimates for SO-induced
lifetimes of plasmon and LO-phonons with the values of the recent Raman
scattering measurements suggests that the effects described above can be
observed experimentally.

\section{Acknowledgements}
We are grateful to Dionys Baeriswyl, Gerd Sch\"{o}n, and Emmanuel Rashba
for valuable discussions. M.P. was supported by the Deutsche 
Forschungsgemeinschaft (DFG). V.G. was supported by the Swiss National Science 
Foundation through Grant No. 20-68047.02.

\appendix

\section{Some technical details of the derivation of $\Pi (q,\omega)$}
\label{techn}

Intermediate steps that lead from from \eq{pilm} to \eq{pregunf}-\eq{funf} are
to express
\begin{widetext}
\beq
g_i (v,z,w,y) &=& \frac{\lambda}{4} \int_0^{2 \pi} d \phi \left[
\left( 1+ \frac{v+ 2 z x}{|{\bf v} + 2 {\bf z}|} \right) \delta
(2 \lambda w - 2 z^2 - 2 v z x + \mu v y - \mu y |{\bf v} + 2 {\bf z}|) \right. \nonumber \\
& & \left. \qquad + \left( 1- \frac{v+ 2 z x}{|{\bf v} + 2 {\bf z}|} \right)
\delta (2 \lambda w  - 2 z^2 - 2 v z x + \mu v y + \mu y |{\bf v} + 2 {\bf z}|) \right], \label{intermed1} \\
f_i (v,z,w,y) &=& \int_0^{2 \pi} \frac{d \phi}{\pi}
\frac{ v z x  + z^2 - \mu y (z x + v) - \lambda w}{( 2 v z x  + 2 z^2 - \mu y v - 2 \lambda w)^2
- y^2 (v^2 + 4 z^2 + 4 v z x)} .
\label{intermed2}
\eeq
\end{widetext}

The change of variables \eq{chanvar} implies
\beq
v &=& -\frac{\tau^2 + y \tau - \lambda w}{\mu \tau}, \label{voft}\\
\frac{\sigma dv}{v} &=& -\mu z \frac{d \tau}{y
\tau}(\lambda_{i+}-\lambda_{i-}),\\
\lambda_{i \sigma} &=& \mu \frac{\tau z -\lambda w (\tau
+y)/z}{\tau^2 + y \tau -\lambda w},\\
\lambda_{i \sigma}^2 - 1 &=& -\frac{\prod_{k=1}^4 (\tau
-\tau_k)}{(\tau^2 + y \tau -\lambda w)^2},\\
v^2 C_i (\lambda_{i\sigma} + \delta_i) &=& \frac{y \tau}{2 z^2}
\frac{(\tau -\tau_3) (\tau -\tau_4)}{\tau^2 + y \tau -\lambda w},
\label{vci}
\eeq
where $\lambda_{i\sigma}$ are defined in \eq{roots}, and $C_i$ and $\tau_k$
are given by \eq{ingred} and \eq{tauk}, respectively.

In this Appendix we also quote table integrals that we use in our 
calculations. They can be either found in Ref.~\onlinecite{gradst} or 
elaborated on its basis. We quote the standard integral \eq{deltint} 
containing the $\delta$-function as well,
\be
\int_0^{2 \pi} d \phi \delta (a+ \cos \phi) = \frac{2}{\sqrt{1 - a^2}} \Theta ( 1 - |a|),
\label{deltint}
\ee
\be
\int_0^{2 \pi} \frac{d \phi}{a+ \cos \phi} = \frac{2 \pi}{\sqrt{a^2 - 1}} 
\Theta ( |a| - 1) {\rm sign} (a),
\label{angint}
\ee
\beq
& & \int_0^{2 \pi} d \phi \frac{x + \delta }{x^2 + \beta x + \gamma}
= \frac{\pi \sqrt{2}}{\sqrt{(1+ \gamma)^2 - \beta^2}} \label{angintcomp}\\
& & \times
\frac{(\delta+1) \sqrt{1 - \beta + \gamma}
+  (\delta -1) \sqrt{1 + \beta + \gamma}}{\sqrt{\sqrt{(1+ \gamma)^2 - \beta^2} 
+ \gamma -1}}, \quad (4 \gamma >\beta^2)
\nonumber
\eeq
\beq
& & \int \frac{d\tau}{\tau-a} \frac{\sqrt{a^2 - 1}}{\sqrt{1 - \tau^2}} 
\label{it1} \\
&=& 2 \arctan  \frac{a-\tau + a \sqrt{1 - \tau^2}}{\tau \sqrt{a^2 - 1}},  
\quad (|\tau|<1<|a|)
\nonumber
\eeq
\beq
& & \int \frac{d \tau}{\tau-a} \frac{\sqrt{1 - a^2}}{\sqrt{1 - \tau^2}}
\label{it2} \\
&=& {\rm ln} \left| \frac{(a-\tau + a \sqrt{1 - \tau^2}
- \tau \sqrt{1 - a^2})^2}{2 a (a -\tau) (1+\sqrt{1 - \tau^2})}\right|, 
\quad (|\tau|, |a|<1).
\nonumber
\eeq
Using \eq{it2} it is easy to establish that
\beq
- \int_{z/y}^{\max (1,z)}\frac{2 \sqrt{z^2 -y^2} d \tau}{z \sqrt{\tau^2 -1}}
\qquad \qquad \label{it21} \\
= -2 \sqrt{1 -\frac{y^2}{z^2}} \left[ \Theta (z-1) {\rm arccosh} z - {\rm arccosh} \frac{z}{y} \right] , \nonumber
\eeq
and
\beq
&-& \int_{0}^{\min (1\mp y,z)}\frac{\sqrt{z^2 -\tau^2} d \tau}{z (\tau \pm y)}
=  1- \Theta (1\mp y -z) \label{it22} \\
& & \times  \left[\sqrt{1 -\frac{y^2}{z^2}}\ln \frac{z+\sqrt{z^2 -y^2}}{y} \pm \frac{\pi y}{2 z} \right] \nonumber \\
&-& \Theta (z-(1\mp y)) \left[\sqrt{1 - \frac{(1 \mp y)^2}{z^2}} \pm \frac{y}{z} \arcsin \frac{1 \mp y}{z} \right. \nonumber \\
&+& \left. 2 \sqrt{1-\frac{y^2}{z^2}} \ln \frac{1 \pm y \sqrt{1 - \frac{(1 \mp y)^2}{z^2}} + (1 \mp y)
\sqrt{1-\frac{y^2}{z^2}}}{\sqrt{2 y \left(1+\sqrt{1 -\frac{(1 \mp y)^2}{z^2}} \right)}}
\right]. \nonumber
\eeq

\section{An alternative representation of $f_{3,4}^{II}$}
\label{compint}

For $4 \gamma_i > \beta_i^2$ ($i=3,4$) the roots of the equation
$x^2+\beta_i x+\gamma_i =0$ are complex. Omitting the index $i$
(in order to avoid confusion with imaginary $i$), we write them in the form ($\sigma =\pm$)
\be
\lambda_{\sigma} = - \frac{\beta}{2}
+i \sigma \sqrt{\gamma - \frac{\beta^2}{4}},
\ee
and note that $\lambda_- = \lambda_+^*$, $|\lambda_{\sigma} |^2 = \gamma$.
The functions $f_{3,4}^{II} \equiv f$ are then represented by
\beq
f &=& - \frac{i C}{2 \pi (\lambda_+ - \lambda_-)}
\sum_{\sigma} {\rm sign}\, (\sigma) \frac{\lambda_{\sigma} + \delta}{\sqrt{\lambda_{\sigma}^2 - 1}} \nonumber \\
& \times &
\oint_{|z|=1} \left( \frac{d z}{z- z_{\sigma +}} -
\frac{d z}{z- z_{\sigma-}} \right), \label{circint}
\eeq
where the secondary roots ($\sigma'=\pm$)
\be
z_{\sigma \sigma'} = \lambda_{\sigma} + \sigma' \sqrt{\lambda_{\sigma}^2 -1}
\ee
obey the equation $z_{\sigma \sigma'}^2 - 2 \lambda_{\sigma} z_{\sigma \sigma'} +1 =0$.
Note the properties $z_{- \sigma'} = z_{+ \sigma'}^*$ and $z_{\sigma+} z_{\sigma-} =1$.
We  also denote $z_{++} = e^{a + ib}$, where $a$ and $b$ are
real.

Checking which of the roots $z_{\sigma \sigma'}$ lie inside the unit circle
$|z|=1$, we find an expression for \eq{circint},
\beq
f  &=& -\frac{C \, {\rm sign} (a)}{\lambda_+ - \lambda_-}
\left[ \frac{\lambda_+ + \delta}{\sqrt{\lambda_+^2 - 1}} -
\frac{\lambda_- + \delta}{\sqrt{\lambda_-^2 - 1}} \right] \nonumber \\
&=& -\frac{C \, {\rm sign} (a)}{{\rm Im} \lambda_+}
{\rm Im}\frac{\lambda_+ + \delta}{\sqrt{\lambda_+^2 - 1}}.
\eeq
One can notice that it resembles its $f^I$-counterpart \eq{fIsigma}, and
therefore a (complex) change of variables similar to \eq{chanvar} would lead
to the form of integrand the same as in \eq{realpi}.
However, the corresponding mapping of the integration contour would assume
that an integration variable $\tau$ runs in the complex plane along the
circle of the radius $\sqrt{w}$. This would probably require an analytic
continuation of the elliptic functions (see Appendix~\ref{elliptic}), which
makes very cumbersome an explicit representation of \eq{repi2}.

\section{Integrals expressed in terms of elliptic functions}
\label{elliptic}

Let us consider the integral
\beq
I =\int \frac{(x-x_a)(x-x_b)}{\sqrt{\prod_{k=1}^4 (x-x_k)}} d x,
\label{intbig}
\eeq
assuming that $x_4< x_3 <x_2 < x_1$. We introduce the notations
$x_{rs}=x_{s}-x_{r}$ and
\be
k^{2}=\frac{x_{32} x_{41}}{x_{31} x_{42}}.
\ee

Let us consider the intervals where the polynomial $\prod_{k=1}^4
(x-x_k)$ is positively defined. For $x < x_4$ and $x_1 < x$ the
integral \eq{intbig} equals
\beq
&I&=\pm \sqrt{\frac{(x-x_1) (x-x_3) (x-x_4)}{x-x_2} } \label{integ14} \\
&+& \frac{1}{\sqrt{x_{31} x_{42}}} [ (2 x_{a2}x_{b2}+ x_{42} x_{21})
F(\varphi,k) -x_{31} x_{42} E(\varphi,k)
\nonumber  \\
&+&  x_{21} (2 x_{a2}+ 2 x_{b2} - x_{32} - x_{42}+x_{21})
\Pi(\varphi,\frac{x_{41}}{x_{42}},k) ], \nonumber
\eeq
where
\be
\varphi = \arcsin \sqrt{\frac{x_{42}}{x_{41}}\frac{x-x_1}{x-x_2}},
\ee
and the sign ``$+$'' has to be chosen for $x_1 <x$, while the sign ``$-$'' has
to be chosen for $x<x_4$.

For $x_3 < x  < x_2$
\beq
&I&=\sqrt{\frac{(x_1 -x) (x_2 - x) (x-x_3)}{x-x_4} }\label{integ23}\\
&+&\frac{1}{\sqrt{x_{31} x_{42}}} [(2 x_{a4}x_{b4}-x_{42}x_{43}) F(\varphi,k) -x_{31} x_{42} E(\varphi,k) \nonumber \\
&+& x_{43} (2 x_{a4}+2 x_{b4} + x_{43}+x_{42}+x_{41})
\Pi(\varphi,\frac{x_{32}}{x_{42}},k) ], \nonumber
\eeq
where
\be
\varphi = \arcsin \sqrt{\frac{x_{42}}{x_{32}}\frac{x-x_{3}}{x-x_{4}}}.
\ee
The elliptic functions
\beq
F(\varphi, k)& =& \int_0^{\varphi}
\frac{d \alpha}{\sqrt{1-k^{2}\sin^{2}\alpha}},\\
E(\varphi, k) &=& \int_0^{\varphi} d \alpha \sqrt{1-k^{2}\sin^{2}\alpha}\ ,\\
\Pi(\varphi,n, k) &=& \int_0^{\varphi}
\frac{d\alpha}{(1-n\sin^{2}\alpha)\sqrt{1-k^{2}\sin^{2}\alpha}},
\eeq
are defined according to Ref.~\onlinecite{gradst}.

\section{Static limit of ${\rm Re} \Pi^{II}$}
\label{statapp}

We start out from the expressions \eq{repi2} and \eq{f3comp}.
For small $w<y^2/4$, we need to consider only the contribution from
the function $f_3^{II}$ specified by $\mu = -$, i.e.,
\be
-\frac{1}{\nu} \lim_{w\to 0} {\rm Re} \Pi^{II} = \lim_{w\to 0} \int_{y-2 \sqrt{w}}^{y+ 2 \sqrt{w}} v f_3^{II} (v,z,w,y) dv .
\label{pi2}
\ee
Let us introduce a new integration variable $\xi =
\frac{v-y}{2\sqrt{w}}$ in \eq{pi2}. Thus, we make the limits of
integration independent of $w$, and at this stage it becomes
possible to exchange the operation $\lim_{w \to 0}$ and the
integration over $\xi$:
\beq
- \frac{1}{\nu} \lim_{w \to 0}  {\rm Re} \Pi^{II} &=& \lim_{w \to 0} \int_{-1}^{1}   d \xi \phi (\xi,z,w,y)\nonumber \\
&=& \int_{-1}^{1} \phi_0 (\xi,z,y) d \xi,
\label{limfin}
\eeq
where
\beq
\phi (\xi,z,w,y) &=& 2 \sqrt{w} (y +2 \xi \sqrt{w}) \nonumber \\
&\times& f_3^{II} (y+2 \xi \sqrt{w},z,w,y) , \\
\phi_0 (\xi,z,y) &=& \lim_{w\to 0} \phi (\xi,z,w,y).
\eeq
We observe that
\beq
 \sqrt{P_-} = \sqrt{z} |z-y| \left[ 1 - \frac{\xi y}{z (z-y)} \sqrt{w} \right. \nonumber \\
+ \left. \frac{z^2 + (z-y)^2 - \xi^2 (y^2 +4 z (z- y))}{2 z^2 (z-y)^2} w + O (w^{3/2})\right], \quad
\eeq
\beq
\sqrt{Q_-} = \sqrt{z} (z+y) \left[ 1  - \frac{\xi y}{z (z+y)} \sqrt{w} \right. \nonumber \\
+ \left. \frac{z^2 + (z+y)^2 - \xi^2 (y^2 + 4 z (z+y))}{2 z^2 (z+y)^2} w + O (w^{3/2})\right]. \quad
\eeq
For $z>y$,
\beq
\sqrt{P_-} +\sqrt{Q_-} &\approx & 2 z \sqrt{z}, \\
\left(\sqrt{P_-} +\sqrt{Q_-}\right)^2 - 4 y^2 z &\approx & 4 z
(z^2-y^2),
\eeq
and therefore $\phi_0 (\xi,z,y) =0$. For $z<y$,
\beq
\sqrt{P_-} +\sqrt{Q_-} &\approx & 2 y \sqrt{z}, \\
\left(\sqrt{P_-} +\sqrt{Q_-}\right)^2 - 4 y^2 z &\approx& \frac{4
y^4 (1-\xi^2)}{z (y^2-z^2)} w,
\eeq
and therefore
\beq
\phi_0 (\xi,z,y) &=& \frac{\sqrt{y^2-z^2}}{z\sqrt{1-\xi^2}}.
\eeq
When inserted in \eq{limfin}, it yields \eq{correct}.

\end{document}